\documentclass[aip,jap,reprint]{revtex4-1}
\usepackage{graphicx}
\usepackage{amsmath}
\usepackage{mathtools}
\usepackage{tabularx}
\usepackage[utf8]{inputenc}
\usepackage[T1]{fontenc}
\usepackage{mathptmx}
 \usepackage{xcolor}
\draft % marks overfull lines with a black rule on the right

\begin{document}

%\preprint{}

\title{Active feedback cooling of a SiN membrane resonator by electrostatic actuation} %Title of paper



\author{A. Borrielli}
\email{borrielli@fbk.eu}
\author{M. Bonaldi}
\author{E. Serra}  \altaffiliation [Also at ]{Istituto Nazionale di Fisica Nucleare (INFN), Trento Institute for Fundamental Physics and Application, 38123 Povo, Trento, Italy}
%\thanks{}
\affiliation{Institute of Materials for Electronics and Magnetism, IMEM-CNR, Trento unit c/o Fondazione Bruno Kessler, Via alla Cascata 56/C, Povo, Trento IT-38123, Italy}

\author{P. M. Sarro}
\author{B. Morana}
\affiliation{Dept. of Microelectronics and Computer Engineering /ECTM/DIMES, Delft University of Technology, Feldmanweg 17, 2628 CT Delft, Netherlands
}

\date{\today}

\begin{abstract}
Feedback-based control techniques are useful tools in precision measurements as they allow to actively shape the mechanical response of high quality factor oscillators used in force detection measurements.
In this paper we implement a feedback technique on a high-stress low-loss SiN membrane resonator, exploiting the charges trapped on the dielectric membrane. A properly delayed feedback force  (dissipative feedback) enables the narrowing of the thermomechanical displacement variance in a similar manner to the cooling  of the normal mechanical mode down to an effective temperature $T_{eff}$. In the experiment here reported we started from room temperature and gradually increasing the feedback gain we were able to cool down the first  normal mode of the resonator to a minimum temperature of about $124 \rm mK$.
This limit is imposed by our experimental set-up and in particular by the the injection of the read-out noise into the feedback. We discuss the implementation details and possible improvements to the technique.
\end{abstract}


\maketitle %\maketitle must follow title, authors, abstract and \pacs

\section{Introduction} \label{Introduction}
Since the pioneering experiments of Coulomb and Cavendish at the end of the sixteenth century up to the refined Micro/Nano Electro-Mechanical Systems (MEMS/NEMS) devices of today, the simple displacement of a mechanical element has played a key role in sensing a wide variety of very weak phenomena with great accuracy.  
As a whole the crucial aspect of these experiments is the ability of converting a weak force to which the mechanical device is subjected  into a displacement $z$ or a frequency shift $\Delta\omega$ that are measurable by electrical or optical high-sensitivity transduction methods.  The detection of single electron  spin \cite{Rugar2004329}, the persistent currents in normal metal rings \cite{Castellanos-Beltran2013}, and   the imprint of quantum phenomena as the force noise associated with the quantized nature of light \cite{Purdy2013801} or the Casimir effect \cite{Bordag20011}, are just a few  of significant achievement that have become possible to day.

In recent decades, thanks in particular to decisive advances in material science and in micro-nanofabrication techniques, it has been possible to design and build mechanical force sensors of extraordinary quality, sensitive in the range of the attonewtons ($10^{-18}N$) at room temperature \cite{Yasumura2000117}, and zeptonewtons ($10^{-21}N$) at cryogenic temperatures \cite{Moser20141007}, under a wide-range of technological platforms as hybrid on-chip structures \cite{Gavartin2012509} carbon nanotubes \cite{Moser2013493} and silicon nitride (SiN) trampoline resonators \cite{Reinhardt2016}.

In principle, force sensitivity is mainly limited by the thermal driving force experienced by the mechanical device, whose power spectral density (PSD) in thermal equilibrium is set by the fluctuation-dissipation theorem (FDT) \cite{Saulson19902437}:
\begin{equation}\label{FSD}
S_{F}=4k_{B}Tm\omega_{0}/Q
\end{equation}
where $k_{B}$ is the Boltzmann's constant, $T$ is the temperature of the environment, $m$ is the  mass and $Q$ is the mechanical quality factor of the resonator. FDT associates thermal Langevin fluctuating forces with the irreversible losses existing in a resonator, quantified by Q in eq. \eqref{FSD}, making it evident that  the fundamental thermomechanical noise floor, and then the Signal-To-Noise ratio (SNR) of a mechanical sensor ,  benefits from a small mass and low mechanical dissipation (high Q). This has justified several decades of research into increasing the quality factor of resonators by reducing the underling dissipations,  leading in recent years into developing devices with unprecedented low mechanical losses \cite{Ghadimi2018764,Tsaturyan2017776}. 

However, in addition to the SNR, the bandwidth of a sensor is important, indeed increasing the Q also decreases the maximum available bandwidth of the system. 
High-Q sensors take a long time (inversely proportional to $Q$) to respond to changes in the external signal, because of the long correlation time of the oscillator motion. The seeking of  the proper trade-off between the general requirements of high SNR (low thermomechanical noise floor) and responsiveness to phenomena that quickly varies in time is therefore a need in the design and manufacture of a sensor. 

\begin{figure}
\includegraphics{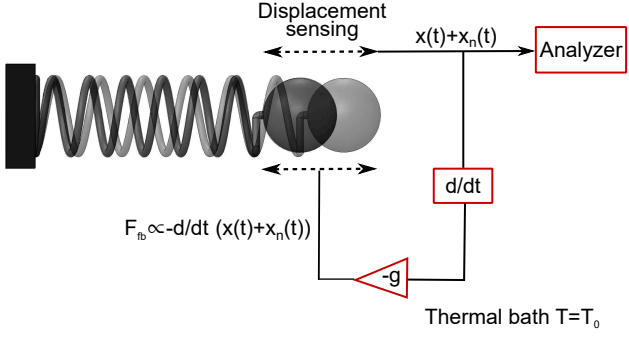}  
\caption{Conceptual scheme of dissipative feedback applied to a damped harmonic oscillator, here represented by a mass $m$ connected to an elastic element of spring constant $k$. The oscillator is real-time actuated by a force proportional to the derivative of the thermomechanical displacement signal. With a negative gain this is equivalent  to an additional viscous damping.}
\label{fig:Fig1}
\end{figure}

A dissipative feedback control of the mechanical system is an effective technique to manage such high quality factors, as those needed to increase the sensitivity, but without placing any restriction on bandwidth.  
In principle, this can be accomplished if the resonator position is externally monitored through a low noise transduction method, phase shifted, and applied back as force; a conceptual representation of dissipative feedback is illustrated in Fig. \ref{fig:Fig1}.
This scheme artificially modifies the mechanical transfer function of the resonator in a similar manner to a change in the effective correlation time of the motion (or in the effective quality factor, $Q_{eff}$), without introducing any  alterations  to the actual dissipation and then to the thermal fluctuations. These techniques are  well established an largely used in a number of fields of physics and engineering; examples can be found in force detection experiments such as those involving ton-scale gravitational wave detectors \cite{Vinante2008}, atomic force microscopes \cite{Mertz19932344} and optomechanical systems \cite{Gavartin2012509}.
Since  dissipative feedback corresponds to a reduction of the thermomechanical mean-squared displacement (displacement variance), many authors refer to these techniques as \textit{cold damping} or \textit{feedback cooling} defining an effective temperature for the resonant mode as follows:
\begin{equation}\label{Qeff}
\frac{Q_{eff}}{Q}\frac{k_{B}T}{m\omega^{2}_{0}}=\frac{k_{B}T_{eff}}{m\omega^{2}_{0}}
\end{equation}
where $T_{eff}=Q_{eff}T/Q$ is the effective temperature of the mode, $Q_{eff}$ the effective quality factor with $Q$ and $T$ the actual values. 

In this paper we present a feedback technique implemented on a SiN membrane-based electromechanical system, and investigate its performances to cool the first mechanical normal mode of the membrane resonator. Starting from room temperature and gradually increasing the feedback gain, we were able to identify a minimum effective temperature $T_{eff}$ of about $124 \rm \ mK$ for the coolest normal mode. This limit is mainly due to the injection of the read-out noise into the feedback that restricts the maximum amount of cooling. 

This kind of membrane-based resonators have been specifically developed to meet the experimental needs of advanced optomechanical setups. Their optical properties are compatible with their use as optomechanical oscillators \cite{Aspelmeyer20141391}, both in Michelson interferometers and in cavity setups \cite{Serra2016}. Moreover, their quality factor remains high in the whole frequency range so they can be used with optimal efficiency, both in single-mode applications, such as optical cooling \cite{Rossi2017}, and in multimode applications such as two-mode squeezing \cite{Patil2015}.  Recently they have been embedded in a “membrane-in-the-middle” setup, allowing to reach a thermal occupation number in the transition region from classical to quantum regime \cite{Chowdhury2019} and to reveal, through the analysis of motional sidebands asymmetry measured by heterodyne detection \cite{Marino2020},  nonclassical properties in the dynamics of macroscopic oscillators \cite{Chowdhury2020}.

Here the feedback was realized by monitoring the thermal motion of the thin membrane ($\approx 100\rm \ nm$) through a high-sensitivity optical interferometric readout and the feedback force was applied by means of an electrode  electrostatically coupled to the trapped charges on the dielectric SiN membrane.

The article is organized as follows: in Sec. \ref{sec:Dissipative feedback theory} we discuss the model for the dissipative feedback and its implication on the performances of a force sensor; in Sec. \ref{sec:experimental} we provide some properties of the membrane-based resonators as well as a detailed description of the experimental realization of the feedback cooling and finally in Sec. \ref{sec:conclusions} we discuss the results.

\section{Dissipative feedback theory} \label{sec:Dissipative feedback theory}
In this paragraph we summarize the standard modeling of the feedback cooling \cite{Poggio2007}. Fig. \ref{fig:Fig1} shows a conceptual diagram of dissipative feedback applied to a damped harmonic oscillator, here represented by a mass $m$, an elastic element  of real spring constant $k$ and an intrinsic dissipation $\gamma_{0}$. The feedback-loop consists into monitoring the thermomechanical motion of the oscillator and actuating it by a force proportional to the  the derivative of the oscillating signal. For a negative gain this corresponds to an additional viscous damping, similarly to the case of a mechanical oscillator subjected to a force proportional to its velocity. 
Since in experiments it is more common  to analyze the mechanical motion in the frequency domain as a noise spectrum, we will examine the effect of  such a feedback force on the spectral features of the thermomechanical motion.

A harmonic oscillator under the condition depicted in Fig. \ref{fig:Fig1} can be described by the Langevin equations written in the frequency domain as:
\begin{align}\label{Langevin}
i\omega x(\omega)~&=v(\omega)\notag\\
i\omega x_{n}(\omega)&=v_{n}(\omega)\nonumber\\
i\omega v(\omega)~&=-\frac{\gamma_{0}}{m}v(\omega)-\omega^{2}_{0}x(\omega)+\frac{1}{m}\xi_{th}(\omega)-\frac{g\gamma_{0}}{m}\left(v(\omega)+v_{n}(\omega)\right)
\end{align}
where $\omega_{0}=\sqrt{k/m}$ is the angular resonance frequency, g is an electronic gain, $x_{n}$ is the read-out noise and  $\xi_{th}(\omega)$ is the frequency component of the random thermal Langevin force with spectral density given by eq. \eqref{FSD}. The last term of eq. \eqref{Langevin} represents the additional viscous force provided by the feedback, and for a more complete modeling of the system it also includes the contribution of the measurement noise on the displacement signal, considering that the measured displacement is $x+x_{n}$.
We can solve eq. \eqref{Langevin} for $x(\omega)$ in terms of the fluctuating force and the measurements noise to obtain:
\begin{equation}\label{DASP}
x(\omega)=\frac{\frac{1}{m}\left(\xi_{th}(\omega)-ig\gamma_{0}\omega x_{n}(\omega)\right)}{\left(\omega^{2}_{0}-\omega^{2}\right)+\frac{1}{m}i(1+g)\gamma_{0}\omega},
\end{equation}
with power spectral density:
\begin{multline}\label{PSD_actual}
S_{x}(\omega)=\frac{1/m^{2}}{\left(\omega^{2}_{0}-\omega^{2}\right)^{2}+\left(1+g\right)^{2}\omega^{2}_{0}\omega^{2}/Q^{2}}S_{F}\\
+\frac{g^{2}\omega^{2}_{0}\omega^{2}/Q^{2}}{\left(\omega^{2}_{0}-\omega^{2}\right)^{2}+\left(1+g\right)^{2}\omega^{2}_{0}\omega^{2}/Q^{2}}S_{x_{n}},
\end{multline}
where $S_{F}$ is the the spectral density of the thermal noise force which depends on the resonator dissipation according to the fluctuation-dissipation theorem (eq.\eqref{FSD}) and $S_{x_{n}}$ is the spectral density of the measurement noise. Considering that our experimental set-up allows a measure of the mechanical dissipation through the quality factor $Q$ (cf. Sec. \ref{sec:experimental}), in writing equation \eqref{PSD_actual} we express the dissipation $\gamma_{0}$ in terms of the oscillator's intrinsic quality factor \cite{nota1} according to $\gamma_{0}=m\omega_{0}/Q$. Now we are able to fix some general features of the feedback scheme presented in Fig. \ref{fig:Fig1} and in the next section we will describe a practical implementation of it.

\subsection{Cold damping} \label{subsec:cold damping}
Fig. \ref{fig:Fig2} shows the normalized thermomechanical displacement power-spectral density (eq. \eqref{PSD_actual}) for $g=0$ (i.e. without feedback) and for increased feedback gains. The dissipative feedback-loop does not change the thermal noise floor that is ultimately set by the intrinsic dissipation of the resonator at the bath temperature. At frequency sufficiently below the resonance ($\omega<<\omega_{0}$) the spectral density is approximately flat with noise floor $\approx\sqrt{4k_{B}T/k\omega_{0}Q}$. The increase of the  feedback gain only modifies the thermal displacement noise at frequencies  close to resonance, and as the most noise power is  concentrated  there, one can observe a gradual decreasing  of the integrated area under the thermomechanical displacement power spectral density (PSD). According to the equipartition theorem we can define the mode temperature of the oscillator as $T_{eff}=k\sigma_{x}^2/k_{B}$ where $\sigma_{x}^2$ represents the variance of the mechanical displacement that is related to the integrated area of the PSD by the relation:

\begin{figure}
\includegraphics{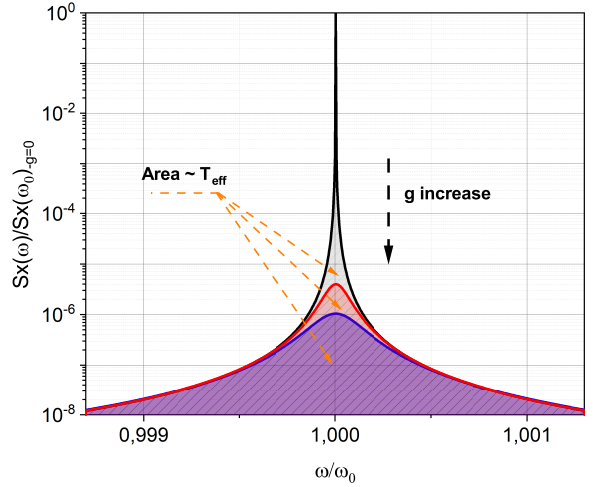}
\caption{Normalized thermomechanical displacement power-spectral density (PSD) in a damped resonator under the effect of a dissipative feedback. Increasing the feedback gain only changes the thermal noise around the resonance, thus observing a gradual decrease of the integrated area that corresponds to a reduction of the effective resonator temperature. PSD was calculated for a resonator with $Q=3.5 \rm \ M$ (condition equivalent to the SiN membrane resonators described in the experimental section) under low gain regime, therefore the first term of \eqref{PSD_actual} was used only.}
\label{fig:Fig2}
\end{figure}

\begin{equation}\label{variance}
\sigma_{x}^2=\frac{1}{2\pi}\int_{0}^\infty S_{x}(\omega)d\omega.
\end{equation}
The consequence of the dissipative feedback is thus a progressive reduction of the effective mode temperature (or equivalently the variance of the thermomechanical noise) which depends on the feedback gain. Using \eqref{PSD_actual} and \eqref{variance} we find:

\begin{equation}\label{Teff}
T_{eff}=\frac{T_{0}}{1+g}+\left(\frac{g^{2}}{1+g}\right)\frac{k\omega_{0}}{4k_{B}Q}S_{x_{n}}
\end{equation}
where $T_{0}$ is the thermal bath temperature. The second term of \eqref{Teff} is due to the injection of the measurement noise (uncorrelated to the Langevin force) into the feedback signal, that imposes a competing heating growing with the feedback gain. The upshot is therefore a minimum achievable temperature  of 
\begin{equation}\label{Tmin}
T_{min}=2T_{0}\frac{\sqrt{1+SNR}-1}{SNR}
\end{equation}
at $g=\sqrt{1+SNR}-1$, where $SNR=4k_{B}QT_{0}/k\omega_{0}S_{x_{n}}$ is the ratio between thermomechanical noise  and readout noise at resonance without feedback.
In the limit of low gain regime ($g<<\sqrt{1+SNR}-1$) the effect of the measurement noise can be neglected because the first term  of \eqref{Teff} is order of magnitudes higher than the second therm and $T_{eff}$ reduces to the familiar form \cite{Kleckner200675} $T_{0}/(1+g)$; on the other hand, at high gains ($g>\sqrt{1+SNR}-1$) the transduction noise sent back to the actuator causes net heating of the vibrational mode. Nevertheless, we stress that the mode temperature $T_{eff}$  isn't a comprehensive property of the system, rather it depicts a single mode of resonance that is pumped out the thermal equilibrium. The bulk temperature in negligibly affected with single mode temperature tuning, because most of the degrees of freedom in the system are not modified by feedback \cite{Miller2018,Hammig2007352}. 
\subsection{Noise squashing} \label{subsec:noise squashing}
A critical aspect of the feedback scheme shown in Fig. \ref{fig:Fig1} can be observed at strong gain, when the motion of the mechanical device is driven by the transduction noise rather than the thermal Langevin force, and consequently it becomes correlated to the noise on the detector. In an in-loop transduction scheme the power spectral density of the measured displacement $x+x_{n}$ can be written as:
\begin{multline}\label{PSD_measured}
S_{x+x_{n}}(\omega)=\frac{1/m^{2}}{\left(\omega^{2}_{0}-\omega^{2}\right)^{2}+\left(1+g\right)^{2}\omega^{2}_{0}\omega^{2}/Q^{2}}S_{F}\\
+\frac{\left(\omega^{2}_{0}-\omega^{2}\right)^{2}+\omega^{2}_{0}\omega^{2}/Q^{2}}{\left(\omega^{2}_{0}-\omega^{2}\right)^{2}+\left(1+g\right)^{2}\omega^{2}_{0}\omega^{2}/Q^{2}}S_{x_{n}}.
\end{multline}
\begin{figure}
\centering
\includegraphics{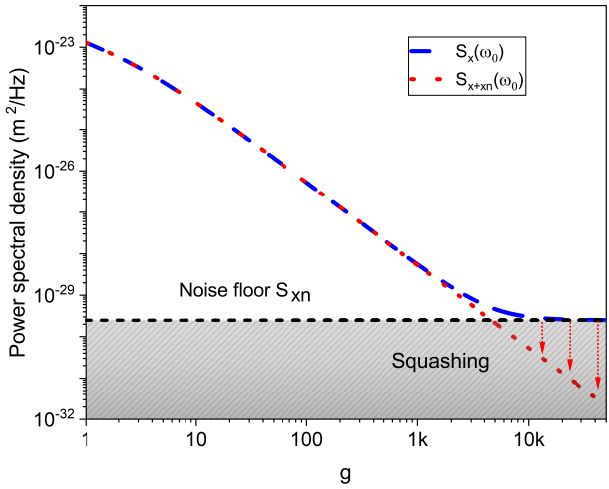}
\caption{Comparison between the  power spectral densities corresponding to the actual displacement \eqref{PSD_actual} and the measured one \eqref{PSD_measured}. Spectra were calculated at  the resonance frequency using the real parameters of the experiment listed in Table \ref{tab:table1}. For large gains the feedback leads to correlations between the resonator displacement and the detector noise, thus changing the shape of the measured power spectrum that is squashed below the detector noise around the resonance.}
\label{fig:Fig3}
\end{figure}
The difference between equation \eqref{PSD_actual} and \eqref{PSD_measured} is schematized in Fig. \ref{fig:Fig3}, where  we calculated the power spectral densities at resonance versus the feedback gain using the parameters listed in Table \ref{tab:table1}, which refer to the SiN membrane resonator and the experimental set-up detailed in the next section. As mentioned earlier the effect of the dissipative feedback is a progressive  drop of the displacement PSD at resonance, and hence the cooling of the mechanical mode. Sufficiently far from the critical gain ($g<<\sqrt{1+SNR}-1 \approx 4600$) the displacement  remains thermal-driven an then uncorrelated from the transduction noise. In this case $S_{x+x_{n}}$ reduces to $S_{x}$ and the mode temperature is proportional to the integrated area between the measured transduction spectra an the transduction noise \cite{Poggio2007}. On the other hand, at higher g  the actual displacement power noise tends to the noise floor $S_{x_{n}}$ but the measured one is reduced  below it. This "squashing" of the noise spectra below the noise of the detector is due  to feedback-induced anticorrelations between the detector noise and the noise-driven displacement, and produces unphysical results if mechanical mode temperature is inferred from the measured power spectra, by simply subtracting the noise floor. Lee et al. in \cite{Lee2010} showed as an independent out-of-loop transduction permits of inferring a mode temperature in good agreement with the prediction of eq. \eqref{Teff}, even above the critical gain.
\subsection{Force sensing resolution} \label{subsec:force resolution}
Linear feedback cooling is a common procedure to increase the bandwidth and reduce the variance of thermal fluctuations in a resonant device. For example, atomic force microscopes (AFM) use a tip hosted in resonant cantilevers as force sensor to map the topography of a surface at the atomic scale, ultimately by sensing the atomic force between the tip and the surface. That resolution requires very low thermal noise floor, and therefore high-Q resonators. Under these conditions the corresponding bandwidth ($\propto 1/Q$) results in a very long response time and therefore an excessive time to scan the surface to be analyzed. Dissipative feedback instead allows to improve the bandwidth and therefore the imaging speed without degrading the SNR \cite{Mertz19932344}. Moreover linear feedback is commonly used to stabilize the tip-surface separation in non-contact configuration AFM, thus avoiding collisions and suppressing frequency drifts \cite{Lee20021154091}.

A more advanced task of oscillator-based force sensors is in detecting a weak signal force $F_{sig}$ with (flat) spectral density well below the thermal background force  $(S_{F}^{sig}<<S_{F}^{th})$, i.e. when the input force is buried on the thermal noise. In ref. \cite{Gavartin2012509} Gavartin et. al experimentally  demonstrated the powerful role of a dissipative feedback protocol in resolving this force against the thermal noise. Indeed,  as long as the energy averaging is chosen as estimator of the force magnitude, the force resolution scales as $\sqrt[4]{\tau_{c}/\tau}$, where $\tau$ is the averaging time and $\tau_{c}$ is the correlation  time of the oscillator.  Thus cold damping represents  a way to effectively improve the convergence of the energy averaging by reducing the correlation time. In that regard, however, we must to point out that:
\begin{itemize}
\item stationary linear feedback doesn't improve the accuracy with which the oscillator position can be determined (i.e. the signal-to-noise ratio), this is because feedback modifies the transfer function of the resonator, and then its response to input excitations,  regardless whether the input is a signal or a background force;
\item as elsewhere discussed \cite{Vinante2013470,Harris2013}, the force estimation process founded on feedback-assisted reduction of the effective resonator time constant isn't necessarily optimal. Indeed,  position measurement recorded with and without feedback are linked by a completely deterministic relation, therefore a proper filtering of the position record without feedback can completely replace the feedback even in the case of nonstationary, non-Gaussian input;
\item appropriate post-processing data filtering requires  accurate knowledge of the susceptibility of the mechanical system that is non trivial especially for micro-optomechanical systems, where the stability of the resonance is affected by several detrimental effects. Stabilization of oscillator parameters and dynamics is thus a crucial issue \cite{Gavartin2013,Pontin2014}.
\end{itemize}

\section{Experimental realization of the feedback cooling} \label{sec:experimental}

\subsection{Silicon-nitride resonators} \label{Silicon-nitride resonators}
The mechanical resonator used throughout the paper for the implementation of the feedback-cooling is a circularly-shaped tensioned SiN membrane, specifically developed to be embedded in advanced optomechanical setups.
Nano-strings or membranes obtained from SiN films have attracted considerable attention due to the possibility of exploiting  the "dilution" of the intrinsic dissipation, usually high in amorphous materials, leading to flexural mechanical modes with very high quality  factors. This effects, first considered in the design of mirror's suspensions in  gravitational wave antennae \cite{Gonzalez2000}, occurs when  SiN layer is produced with  residual stress which is tensile and high enough to push  the device to a regime where the mechanical behavior is governed by the internal stress of the layer and the flexural rigidity can be neglected \cite{Schmid2011}. Dissipation dilution effects can be observed in very thin strings or membranes with thicknesses smaller than $100 ~\rm nm$ and residual stress starting from about $1~ \rm GPa$. The use of these structures combined with efficient solutions to isolate them from their support has allowed realizing resonators with exceptionally high Q factors \cite{Tsaturyan2017776,Ghadimi2018764}.

In the field of optomechanics, systems based on a SiN membrane oscillator have shown for the first time the mechanical effect of the quantum noise in the light \cite{Purdy2013801} and one of the first observations of pondero-motive light squeezing \cite{Purdy2014}. In many cases, the oscillators consist of commercially available free-standing SiN membranes having  residual tensile stress of $1~\rm GPa$. However, in this case, the mechanical quality factors of many millions in principle obtainable thanks to the high tensile stress, cannot be achieved due to the mechanical losses. These are strongly dependent on the mounting, especially for the low frequency modes, and are the cause of scattered Q-factors based-on  the modal form \cite{Wilson2009}.

Recently we proposed a way to overcome the limitations of  commercial membranes by addressing the issue of mechanical losses employing a coupled oscillators model \cite{Borrielli2016}, where the vibrations of the membrane are considered along with those of the silicon chip. This approach resulted in the design and realization of a "loss shield" structure on which the membrane is embedded (Fig. \ref{fig:Fig4}a ). In these devices almost all of the vibrations of the membrane have a high quality factor and reach the limit set by the intrinsic dissipation.
The production of these devices has required the development of a specific manufacturing process\cite{Serra20181193} based on MEMS bulk micro-machining by Deep-Reaction Ion Etching (DRIE) and through  two-side wafer processing.

The SiN employed for the realization of the circularly-shaped membranes was obtained by means of LPCVD (Low-Pressure Chemical Vapor Deposition) using $DCS$ (Dichlorosilane) and $NH_{3}$ (Ammonia) as gas precursors. The gas flows were tuned to obtain a SiN layer with a composition close to stoichiometry ($Si_{3}N_{4}$) and a resulting residual tensile stress close to $1~\rm GPa$. The deposition time was tuned to obtain a SiN layer with a thickness of $100 ~\rm nm$. Although a layer with higher stress and lower thickness would have allowed to improve the mechanical performance, owing to a greater dilution effect and less dissipation,  the above values were selected as they offered a good trade-off. More specifically such layers allowed high mechanical quality factors (Q), low absorption of the film at the laser wavelength used in our optomechanical setups ($1064 ~\rm nm$), and the required robustness for the achievement of large area membranes capable of withstanding all the necessary fabrication and subsequent cleaning and handling steps. The residual tensile stress, measured by the wafer curvature method and confirmed by the resonance frequencies of the device, was slightly lower ($0.83~ \rm Gpa$) than the target value of $1 ~\rm GPa$. Most likely this was due to  not perfect control of the deposition pressure of the employed LPCVD system. The film thickness and the refractive index were determined by variable-angle spectroscopic ellipsometry. The thickness resulted in compliance with the nominal value within 5\%. The refractive index of the film at $1064 ~\rm nm$ was instead $\rm n=1.993$ with the error below 1\%, this is slightly different from that of a perfectly stoichiometric SiN film, which is $\rm n=2$ at $1064 ~\rm nm$. This was expected as the gas flow ratio ($DCS/NH_{3}$) specifically set to obtain residual tensile stress at the target value was slightly lower compared to the one used for the deposition of $Si_{3}N_{4}$ in the employed LPCVD system.

\subsection{Experimental set-up} \label{Experimental set-up}
\begin{figure}
\centering
\includegraphics{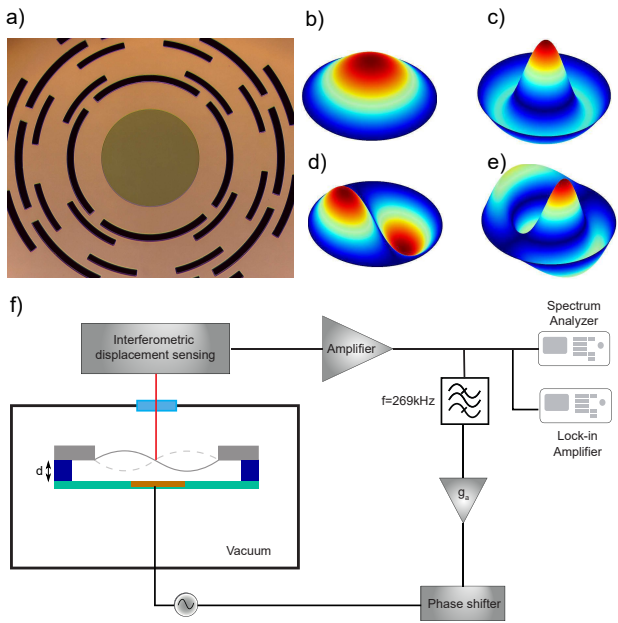}
\caption{a) Optical microscope picture of the circular membrane, with diameter $1.55 \rm \ mm$ and thickness $100 \rm \ nm$. The membrane is mechanically supported by silicon structure acting as "loss shield" \cite{Borrielli2016}. b-e) modal shapes of the first four membrane modes, resonating at frequencies between $269$ and $621 \rm \ kHz$. f) Scheme of the feedback loop realized by an interferometric displacement sensing and an electrostatic actuation of the membrane. }
\label{fig:Fig4}
\end{figure}

The experimental set-up used for the implementation of the feedback-cooling is shown in Fig. \ref{fig:Fig4}f and includes  a displacement  sensing section   and a feedback-controlled actuation section. The first one is described in details elsewhere \cite{Serra20181193} and basically consists into a polarization-sensitive Michelson interferometer with displacement sensitivity of about $10^{-30} \rm \ m^{2}/Hz$. Here it is employed to sense the thermomecanical noise of the  tensioned membranes.  Circular tensioned membranes have normal modes whose frequencies are given by the expression $f_{mn}=\frac{1}{2\pi}\sqrt{\frac{\sigma}{\rho}}\frac{1}{R}\alpha_{mn}$, where $\sigma$ is the stress, $\rho$ the density, $R$ the radius of the membrane and $\alpha_{mn}$ is the n-th root of the first kind Bessel function of order $m$. Figures \ref{fig:Fig4}b-e show the modal shapes of the first four modes (with indexes (0,1) (0,2) (1,1) and (1,2)). In the present work the feedback cooling was realized on the first mode (0,1) with frequency about $269 \rm \ kHz$.

The feedback force was applied by an electrostatic actuation scheme as depicted in Fig. \ref{fig:Fig4}f. A circular electrode realized on a PCB board is firmly clamped to the silicon chip supporting the membrane and it is held at distance $d \approx 300 \rm \  \mu  m$ from it by calibrated spacers. The actuation of the membrane is controlled by the voltage $V$ applied to the electrode through an external circuit.
In such a configuration several phenomena could contribute to  impart forces on the membrane, which can be either electrostatic, due to the fixed charges ordinarily accumulated on the surface of dielectric medium as SiN, or dielectric in case a static voltage $V_{dc}$ polarizes the dielectric device which in turn is subjected to an attractive force. In both cases the forces can be modulated at high frequency and  several groups reported effective methodologies of electrostatic transduction/activation, integrated on MEMS/NEMS devices \cite{Unterreithmeier20091001,Buters2017,Schmid2014}. 

We characterized the forces, imparted on the resonator by the actuation system of Fig. \ref{fig:Fig4}f, by applying both a bias voltage $V_{dc}$ and a weak modulation voltage $V_{ac}$ to the electrode and sending the displacement signal to a lock-in amplifier (HF2LI Zurich Instruments). With this scheme we drove the mechanical resonance of the membrane performing three different sets of tests.
As first we have varied $V_{dc}$ in the range $0-10 \rm \ V$ and kept $V_{ac}$ at a fixed value of  $5 \rm \ mV$ with a modulation frequency equal to the mechanical resonance frequency of the first normal mode of the membrane ($f_{0}=269 \rm \ kHz$). The amplitude of the mechanical oscillation recorded by the lock-in was found to be  independent from the value of $V_{dc}$, which excludes effects  of dielectric forces that would instead be dependent on the applied bias voltage. 
Secondly we set $V_{dc}$ at zero  and varied $V_{ac}$ in the range $1-10 \rm \ mV$ at the resonance frequency. In this case the mechanical response of the membrane was linear with $V_{ac}$. 
Finally, in order to verify the presence or absence of force components proportional to $V_{ac}^{2}$,  we applied a variable voltage $V_{ac}$ at half the mechanical resonance frequency $f_{0}/2$, noting the absence of any effect on the resonant mode. 
These three tests allow us to conclude that in this case the membrane actuation is linear and it occurs by purely electrostatic effects due to the accumulation of trapped charges on the dielectric membrane.The charging of SiN layers is a well-known phenomenon due to the capture of charges by trapping centers originating from a specific structural defect in amorphous silicon nitride \cite{Krick19888226}.

\begin{table} 
\caption {Parameters characterizing the fundamental mode $(0,1)$ considered for the feedback, as well as other relevant parameters concerning both the membrane and measuring system. The  modal mass $m$ is extracted as fit parameter from Fig. \ref{fig:Fig5}A with an error of $\pm 10 \%$. The other quantities are obtained from independent measurements with an error below $2 \%$.}
\label{tab:table1}
\begin{ruledtabular}
\begin{tabular}{ccccccc}

m &$\rho$ &$\sigma$  &f      & R       &Q           &$S_{x_{n}}$              \\
\hline

($\rm \mu g$)  & ($\rm \frac{g}{cm^{3}}$)   & ($\rm Gpa$)   & ($\rm kHz$)  & ($\rm mm$)    &             & ($\rm \frac{m^{2}}{Hz}$)      \\
\hline

0.21     & 2.8                  & 0.83     & 269    & 0.77     & $3.5x10^{6}$   & $2.3x10^{-30}$            \\
\end{tabular}
\end{ruledtabular}
\end{table} 

\begin{figure*}
\centering
\includegraphics{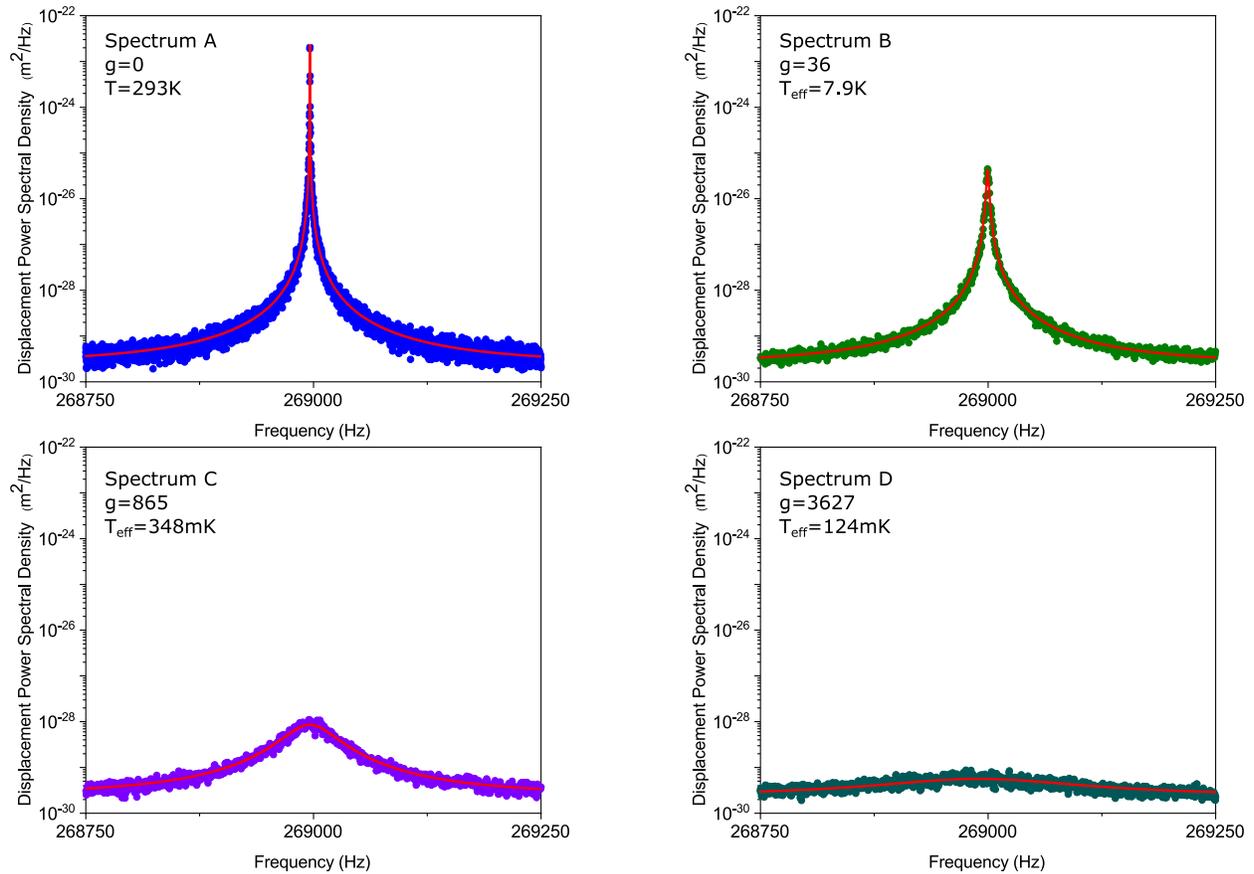}
\caption{Cold damping  observed on the measured spectral densities of the first resonant mode. Spectrum A represents the oscillator in thermal equilibrium ($g=0$) at room temperature ($293K$) while B-D are obtained for different feedback gains. The effective mode temperatures reported in spectra B-D are inferred by fitting the experimental spectra to equation \eqref{PSD_measured} (red lines) and by substituting the extracted parameters to equation \eqref{Teff}. The inferred gains and mode temperatures are given with an error of $\pm 10 \%$ }
\label{fig:Fig5}
\end{figure*}

Another relevant point to check  for the feedback experiment is the stability of the electric charge. In fact the overall feedback gain $g$ is set by a combination of an adjustable gain amplifier $g_{a}$ and of a transduction gain $g_{t}$, so that $g=g_{a}g_{t}$, where $g_{t}$  depends on the electric charge.
We have verified that the electric charge remains constant over time (at least in the period of observation which was 1 day) if the experimental set-up remains under vacuum, therefore $g_{t}$ can be considered a constant parameter in the feedback experiment, and as discussed in the next section it is derived as fit parameter. 

The feedback loop consists of an amplifier, a bandpass filter and phase shifter (Fig. \ref{fig:Fig4}f). 
The need for the filters is determined by the presence of a variety of membrane oscillation modes, with frequencies and modal shapes different from the fundamental mode considered in this work. To realize a dissipative feedback and avoid self-oscillation of the mechanical system, the feedback force should be proportional and opposite to the speed of the displacement, therefore having a time lag corresponding to $-90^\circ$, at all frequencies corresponding to the modal forms. This is very difficult for two reasons:
\begin{enumerate}
	\item the phase shift in an analog circuit is usually introduced with a combination of bandpass filters, therefore being frequency-dependent it cannot keep a constant value over a wide band;
	\item  higher order modes have radial and/or circumferential nodal lines, so the overall displacement of the membrane is the overlapping of modal shapes that can move in opposite directions, even if at different frequencies. Therefore, moving from one mode to another may require not only an adjustment of the phase, but also a change of sign of implementation. The problem is practically insurmountable in the case of mechanical doubles, which move with opposite phases and can have only a few tens of Hz difference.
\end{enumerate}

Since a force applied with the wrong phase can activate self-oscillations of the system, electromechanical feedback is only possible on  normal modes with enough separate frequencies so that the bandpass filter makes negligible the force applied to the mechanical modes nearby in frequency.
In addition to the fundamental mode, in a round membrane the possible candidates are all modes with axial symmetry, therefore without radial nodal lines. Moreover, since the modal density increases strongly with frequency, the possible candidates are to be searched among the modes with lower frequency.
To allow the use of a high feedback gain in the cooling experiment, we implemented a cascade of multi-polar LC filters to a total order of 7. 
To reduce the cross talk between input and output of the filter system, the filter cascade was built into two shielded boxes, separated by a variable gain amplifier $g_{a}$ with an adjustable phase shifter. 

Finally the whole experimental set-up is accommodated in a vibration isolated vacuum chamber, at a base pressure of $10^{-6} \rm \ mbar$, to suppress gas damping effects on the resonator properties. 
\subsection{Experimental observation of cold damping} \label{subsec:experimental observation}
As a first step we have characterized the resonant  mode of the membrane in terms of resonance frequency and mechanical quality factor at room temperature and without feedback. The quality factor was  evaluated by the ring-down method, i.e. the resonator was first forced at the resonance frequency by means of a piezoelectric crystal fixed to the sample holder, then the drive voltage was removed and the vibration recorded during the decay.  The decay time of the mode gives a direct estimation of the Q-factor \cite{Serra20181193}. The measured Q was $3.5x10^{6}$, appreciably lower than the typical value \cite{Borrielli2016} for this type of membrane ( $\approx10^{7}$), probably due to some contamination of the membrane surfaces  during electrode clamping.

Secondly we used a spectrum analyzer to measure the displacement spectral density of the membrane around the resonance mode of interest with the system disconnected from the feedback $(g=0)$ (Fig. \ref{fig:Fig5}A). This first spectrum allowed to calibrate the effective modal mass by fitting the  spectral density data to the equation \eqref{PSD_measured}  where  all other parameters have been measured independently. Note that for $g=0$ eq. \eqref{PSD_measured}  reduces to the more familiar form given in  \cite{Saulson19902437}. Modal mass was found to be of $0.21 \pm 0.02 \rm \ \mu g$, in agreement  with the theoretical values \cite{Serra2016}. In Table \ref{tab:table1} we summarize the parameters characterizing the fundamental mode $(0,1)$ considered for the feedback, as well as other relevant parameters concerning both the membrane and the measuring system.
\begin{figure}
\centering
\includegraphics{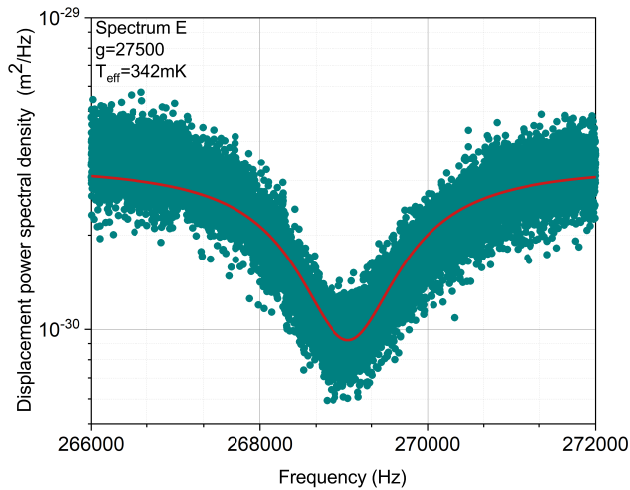}
\caption{Distortion of the displacement noise caused by feedback-induced correlations with the detector noise at high gain regime. At this gain level the feedback heats the resonator rather than cool it, and the effective temperature is inferred by fitting the data to \eqref{PSD_measured}  and using \eqref{Teff} . Alternatively a second detector (out of loop) could be used to measure the resonator motion \cite{Lee2010}.}
\label{fig:Fig6}
\end{figure}

Finally  we turned on the feedback and we progressively increased the feedback gain by tuning  the amplifier gain $g_{a}$. In Fig. \ref{fig:Fig5}B-D we show a set of displacement spectral densities  obtained for three different values of $g_{a}$. The effective mode temperatures shown in Fig. \ref{fig:Fig5} are inferred by fitting the experimental spectra to equation \eqref{PSD_measured}  and by substituting the extracted parameters to equation \eqref{Teff} . The fits comprise three free parameters $\omega_{0}$, $S_{x_{n}}$ and  the overall feedback gain $g$, that in the three examples shown is $36$, $865$ and $3627$ respectively, with an error of $\pm 10\%$. As discussed in the previous section, the feedback force applied with the the right phase results in a cooling of the mechanical mode and the effective modal temperature drops from room temperature at $g=0$  down to $T_{eff}=124 \pm 12 \rm \ mK$ at $g=3627$, which is the minimum temperature reached in this experiment.

Note that if the gain remains well below the critical value defined in section \ref{subsec:cold damping}, which in this specific case is $\approx 4600$, the observed thermal noise spectra are orders of magnitude higher than  the measurement noise, witch  implies that the mode temperature could be well determined also by the area between the observed spectra and the noise floor.  Such a condition occurs for spectra A-C of Fig. \ref{fig:Fig5}, but it is no longer valid  for spectrum D where the gain is close to the critical one and  only eq. \eqref{Teff}  allows an accurate temperature estimation.

A further increase of the gain causes a fall of the observed thermal noise below to the white noise floor marking the noise limit of the transduction system, this "squashing" is shown in Fig. \ref{fig:Fig6} for $g=27500$. As discussed in sec. \ref{subsec:noise squashing}, at this gain level the feedback loop has the opposite effect to the desired, exciting the resonator rather than damping it, as result an increase of the mode temperature is observed.

A suggestive  evolution of this cold-damping scheme could be to replicate  n times the pair bandpass filter/phase shifter of Fig. \ref{fig:Fig4}f, so as to cool all the resonant modes with  frequencies $f_{1}...f_{n}$ included in a  wide range. This multimode cold-damping could lead to the partial refrigeration of the mechanical resonator, in contrast to the single mode cooling that leaves the overall temperature of the object largely unaltered. Recently, the possibility of efficiently  extracting thermal energy from many vibrational modes via cold-damping feedback,  has been predicted under the condition of frequency-resolved resonators \cite{Sommer2019}.  This condition, however,  results virtually inapplicable to the membrane resonator used throughout the article, mostly due to the mechanical doublets occurring for non-axisymmetric mechanical modes (with index $m\neq 0$) that are separated by few tens of Hz (cf. section \ref{Experimental set-up})

\section{Discussion and conclusions} \label{sec:conclusions}

In this work we have detailed the realization of a strong feedback cooling on the first normal mode of a tensioned SiN membrane. We exploited a viscous feedback electrostatic force that increases the damping of the resonator without adding thermal fluctuations.  Starting from room temperature we were able to measure a 3 orders of magnitude reduction of the effective modal temperature, reaching the limits imposed by the read-out. The scheme here adopted takes advantage of the trapped charges on the large-area dielectric SiN membrane and represents an easy  system of  manipulation and  displacement control for that kind of  devices.
It is simple to achieve by not requiring high DC voltage to actuate the device, with noticeable benefits in terms of reduction in the complexity of the experimental set-up, and  the actuation electrodes could be easily built-in on the device chip at  the microfabrication stage. Also it is in principle compatible with more complex optomechanical setups as for example the cavity detuning control in a membrane-in-the-middle optomechanical configuration \cite{Thompson200872}.

In that regard it is worth pointing out that micromechanical resonators as SiN membranes coupled to optical technologies represent a promising platform to test  quantum  limits of resonant sensors or more in general quantum non-classical behavior in macroscopic  objects \cite{Chowdhury2020,Marino2020}. An important prerequisite for approaching the quantum realm in a resonant device is that it is into its quantum ground state. At any non-zero temperature there is always a finite probability  to find the resonator in an excited state, and the average thermal occupation is \cite{Poot2012273} $\left\langle n\right\rangle  \approx k_{B} T/\hbar \omega_{0}-1/2$. When  $\left\langle n\right\rangle =1$ the probability finding the resonator  in the ground state is $50\%$, this means that  $\left\langle n\right\rangle \leq 1$ indicates that the resonator is into its ground state most of the time. This is  achievable with ultra-high frequency resonators ($f\geq \rm GHz$) in a dilution refrigerator ($T\approx 50 \rm \ mK$), but lower-frequency resonators and higher environment temperature require cooling techniques as the active feedback cooling here reported or sideband cooling \cite{Marquardt2009}.

\begin{figure}
\centering
\includegraphics{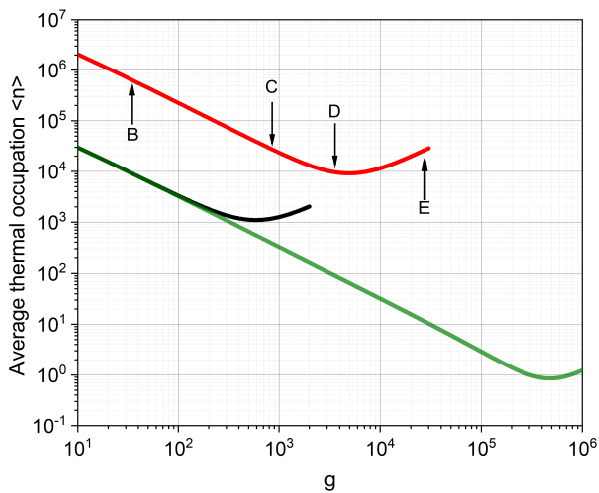}
\caption{Evaluation of the average thermal occupation obtainable by active feedback cooling. An upgrade to liquid helium temperature of the experimental set-up would allow to improve the results reported in the paper (red line) up to $\left\langle n\right\rangle \approx 1000$ (black line).  Cooling the membrane to its ground state ($\left\langle n\right\rangle \leq 1$) would requires  a measurement noise reduced to $10^{-35} \rm \ m^{2}/Hz$}
\label{fig:Fig7}
\end{figure}

Although in our  experiment at room temperature we have reached a modest average occupation number $\left\langle n\right\rangle \approx 10^{4}$,  this figure could be greatly reduced by improving the parameters limiting the maximum available amount of cooling, i.e  the thermal bath temperature $T_{0}$, the quality factor of the resonant mode and especially the  read-out noise injected into the feedback.
As an example in Fig. \ref{fig:Fig7} we compare the results obtained in the experiment presented in the paper with those expected in a state-of-the-art experimental set-up.
The red line shows the trend of the average thermal occupation number as a function of the feedback gain, calculated on the basis of \eqref{Teff}  and by using the experimental parameters of Table \ref{tab:table1}. In the curve we marked the positions corresponding to the spectra B,C,D and E reported in figures \ref{fig:Fig5} and \ref{fig:Fig6}. The black line is instead an evaluation of the thermal occupation number expected  if the same experiment would be conducted at $4.2 \rm \ K$, the base temperature of standard cryogenic systems based on liquid helium. 
Finally, with the green curve we show how it is possible to obtain a near ground state final occupation number $\approx 0.8$ for a resonators with quality factor of $10 \rm \ M$, that is the standard for the SiN membrane considered here,  by reducing the measurement noise up to $10^{-35} \rm \ m^{2}/Hz$ which can be attained in readout configurations based on optomechanical cavities \cite{Purdy2014}.

\begin{acknowledgments}
This research was performed within the Project QuaSeRT funded by the QuantERA ERA-NET Cofund in Quantum Technologies
implemented within the European Union’s Horizon 2020 Programme. The research has been partially supported by INFN
(HUMOR project).\\
The data that support the findings of this study are available
from the corresponding author upon reasonable request.
\end{acknowledgments} 
\bibliography{source_bib}

\begin{thebibliography}{46}%
\makeatletter
\providecommand \@ifxundefined [1]{%
 \@ifx{#1\undefined}
}%
\providecommand \@ifnum [1]{%
 \ifnum #1\expandafter \@firstoftwo
 \else \expandafter \@secondoftwo
 \fi
}%
\providecommand \@ifx [1]{%
 \ifx #1\expandafter \@firstoftwo
 \else \expandafter \@secondoftwo
 \fi
}%
\providecommand \natexlab [1]{#1}%
\providecommand \enquote  [1]{``#1''}%
\providecommand \bibnamefont  [1]{#1}%
\providecommand \bibfnamefont [1]{#1}%
\providecommand \citenamefont [1]{#1}%
\providecommand \href@noop [0]{\@secondoftwo}%
\providecommand \href [0]{\begingroup \@sanitize@url \@href}%
\providecommand \@href[1]{\@@startlink{#1}\@@href}%
\providecommand \@@href[1]{\endgroup#1\@@endlink}%
\providecommand \@sanitize@url [0]{\catcode `\\12\catcode `\$12\catcode
  `\&12\catcode `\#12\catcode `\^12\catcode `\_12\catcode `\%12\relax}%
\providecommand \@@startlink[1]{}%
\providecommand \@@endlink[0]{}%
\providecommand \url  [0]{\begingroup\@sanitize@url \@url }%
\providecommand \@url [1]{\endgroup\@href {#1}{\urlprefix }}%
\providecommand \urlprefix  [0]{URL }%
\providecommand \Eprint [0]{\href }%
\providecommand \doibase [0]{http://dx.doi.org/}%
\providecommand \selectlanguage [0]{\@gobble}%
\providecommand \bibinfo  [0]{\@secondoftwo}%
\providecommand \bibfield  [0]{\@secondoftwo}%
\providecommand \translation [1]{[#1]}%
\providecommand \BibitemOpen [0]{}%
\providecommand \bibitemStop [0]{}%
\providecommand \bibitemNoStop [0]{.\EOS\space}%
\providecommand \EOS [0]{\spacefactor3000\relax}%
\providecommand \BibitemShut  [1]{\csname bibitem#1\endcsname}%
\let\auto@bib@innerbib\@empty
%</preamble>
\bibitem [{\citenamefont {Rugar}\ \emph {et~al.}(2004)\citenamefont {Rugar},
  \citenamefont {Budakian}, \citenamefont {Mamin},\ and\ \citenamefont
  {Chui}}]{Rugar2004329}%
  \BibitemOpen
  \bibfield  {author} {\bibinfo {author} {\bibfnamefont {D.}~\bibnamefont
  {Rugar}}, \bibinfo {author} {\bibfnamefont {R.}~\bibnamefont {Budakian}},
  \bibinfo {author} {\bibfnamefont {H.}~\bibnamefont {Mamin}}, \ and\ \bibinfo
  {author} {\bibfnamefont {B.}~\bibnamefont {Chui}},\ }\href {\doibase
  10.1038/nature02658} {\bibfield  {journal} {\bibinfo  {journal} {Nature}\
  }\textbf {\bibinfo {volume} {430}},\ \bibinfo {pages} {329} (\bibinfo {year}
  {2004})}\BibitemShut {NoStop}%
\bibitem [{\citenamefont {Castellanos-Beltran}\ \emph
  {et~al.}(2013)\citenamefont {Castellanos-Beltran}, \citenamefont {Ngo},
  \citenamefont {Shanks}, \citenamefont {Jayich},\ and\ \citenamefont
  {Harris}}]{Castellanos-Beltran2013}%
  \BibitemOpen
  \bibfield  {author} {\bibinfo {author} {\bibfnamefont {M.}~\bibnamefont
  {Castellanos-Beltran}}, \bibinfo {author} {\bibfnamefont {D.}~\bibnamefont
  {Ngo}}, \bibinfo {author} {\bibfnamefont {W.}~\bibnamefont {Shanks}},
  \bibinfo {author} {\bibfnamefont {A.}~\bibnamefont {Jayich}}, \ and\ \bibinfo
  {author} {\bibfnamefont {J.}~\bibnamefont {Harris}},\ }\href {\doibase
  10.1103/PhysRevLett.110.156801} {\bibfield  {journal} {\bibinfo  {journal}
  {Physical Review Letters}\ }\textbf {\bibinfo {volume} {110}} (\bibinfo
  {year} {2013}),\ 10.1103/PhysRevLett.110.156801}\BibitemShut {NoStop}%
\bibitem [{\citenamefont {Purdy}, \citenamefont {Peterson},\ and\ \citenamefont
  {Regal}(2013)}]{Purdy2013801}%
  \BibitemOpen
  \bibfield  {author} {\bibinfo {author} {\bibfnamefont {T.}~\bibnamefont
  {Purdy}}, \bibinfo {author} {\bibfnamefont {R.}~\bibnamefont {Peterson}}, \
  and\ \bibinfo {author} {\bibfnamefont {C.}~\bibnamefont {Regal}},\ }\href
  {\doibase 10.1126/science.1231282} {\bibfield  {journal} {\bibinfo  {journal}
  {Science}\ }\textbf {\bibinfo {volume} {339}},\ \bibinfo {pages} {801}
  (\bibinfo {year} {2013})}\BibitemShut {NoStop}%
\bibitem [{\citenamefont {Bordag}, \citenamefont {Mohideen},\ and\
  \citenamefont {Mostepanenko}(2001)}]{Bordag20011}%
  \BibitemOpen
  \bibfield  {author} {\bibinfo {author} {\bibfnamefont {M.}~\bibnamefont
  {Bordag}}, \bibinfo {author} {\bibfnamefont {U.}~\bibnamefont {Mohideen}}, \
  and\ \bibinfo {author} {\bibfnamefont {V.}~\bibnamefont {Mostepanenko}},\
  }\href {\doibase 10.1016/S0370-1573(01)00015-1} {\bibfield  {journal}
  {\bibinfo  {journal} {Physics Report}\ }\textbf {\bibinfo {volume} {353}},\
  \bibinfo {pages} {1} (\bibinfo {year} {2001})}\BibitemShut {NoStop}%
\bibitem [{\citenamefont {Yasumura}\ \emph {et~al.}(2000)\citenamefont
  {Yasumura}, \citenamefont {Stowe}, \citenamefont {Chow}, \citenamefont
  {Pfafman}, \citenamefont {Kenny}, \citenamefont {Stipe},\ and\ \citenamefont
  {Rugar}}]{Yasumura2000117}%
  \BibitemOpen
  \bibfield  {author} {\bibinfo {author} {\bibfnamefont {K.}~\bibnamefont
  {Yasumura}}, \bibinfo {author} {\bibfnamefont {T.}~\bibnamefont {Stowe}},
  \bibinfo {author} {\bibfnamefont {E.}~\bibnamefont {Chow}}, \bibinfo {author}
  {\bibfnamefont {T.}~\bibnamefont {Pfafman}}, \bibinfo {author} {\bibfnamefont
  {T.}~\bibnamefont {Kenny}}, \bibinfo {author} {\bibfnamefont
  {B.}~\bibnamefont {Stipe}}, \ and\ \bibinfo {author} {\bibfnamefont
  {D.}~\bibnamefont {Rugar}},\ }\href {\doibase 10.1109/84.825786} {\bibfield
  {journal} {\bibinfo  {journal} {Journal of Microelectromechanical Systems}\
  }\textbf {\bibinfo {volume} {9}},\ \bibinfo {pages} {117} (\bibinfo {year}
  {2000})}\BibitemShut {NoStop}%
\bibitem [{\citenamefont {Moser}\ \emph {et~al.}(2014)\citenamefont {Moser},
  \citenamefont {Eichler}, \citenamefont {Güttinger}, \citenamefont {Dykman},\
  and\ \citenamefont {Bachtold}}]{Moser20141007}%
  \BibitemOpen
  \bibfield  {author} {\bibinfo {author} {\bibfnamefont {J.}~\bibnamefont
  {Moser}}, \bibinfo {author} {\bibfnamefont {A.}~\bibnamefont {Eichler}},
  \bibinfo {author} {\bibfnamefont {J.}~\bibnamefont {Güttinger}}, \bibinfo
  {author} {\bibfnamefont {M.}~\bibnamefont {Dykman}}, \ and\ \bibinfo {author}
  {\bibfnamefont {A.}~\bibnamefont {Bachtold}},\ }\href {\doibase
  10.1038/nnano.2014.234} {\bibfield  {journal} {\bibinfo  {journal} {Nature
  Nanotechnology}\ }\textbf {\bibinfo {volume} {9}},\ \bibinfo {pages} {1007}
  (\bibinfo {year} {2014})}\BibitemShut {NoStop}%
\bibitem [{\citenamefont {Gavartin}, \citenamefont {Verlot},\ and\
  \citenamefont {Kippenberg}(2012)}]{Gavartin2012509}%
  \BibitemOpen
  \bibfield  {author} {\bibinfo {author} {\bibfnamefont {E.}~\bibnamefont
  {Gavartin}}, \bibinfo {author} {\bibfnamefont {P.}~\bibnamefont {Verlot}}, \
  and\ \bibinfo {author} {\bibfnamefont {T.}~\bibnamefont {Kippenberg}},\
  }\href {\doibase 10.1038/nnano.2012.97} {\bibfield  {journal} {\bibinfo
  {journal} {Nature Nanotechnology}\ }\textbf {\bibinfo {volume} {7}},\
  \bibinfo {pages} {509} (\bibinfo {year} {2012})}\BibitemShut {NoStop}%
\bibitem [{\citenamefont {Moser}\ \emph {et~al.}(2013)\citenamefont {Moser},
  \citenamefont {Güttinger}, \citenamefont {Eichler}, \citenamefont
  {Esplandiu}, \citenamefont {Liu}, \citenamefont {Dykman},\ and\ \citenamefont
  {Bachtold}}]{Moser2013493}%
  \BibitemOpen
  \bibfield  {author} {\bibinfo {author} {\bibfnamefont {J.}~\bibnamefont
  {Moser}}, \bibinfo {author} {\bibfnamefont {J.}~\bibnamefont {Güttinger}},
  \bibinfo {author} {\bibfnamefont {A.}~\bibnamefont {Eichler}}, \bibinfo
  {author} {\bibfnamefont {M.}~\bibnamefont {Esplandiu}}, \bibinfo {author}
  {\bibfnamefont {D.}~\bibnamefont {Liu}}, \bibinfo {author} {\bibfnamefont
  {M.}~\bibnamefont {Dykman}}, \ and\ \bibinfo {author} {\bibfnamefont
  {A.}~\bibnamefont {Bachtold}},\ }\href {\doibase 10.1038/nnano.2013.97}
  {\bibfield  {journal} {\bibinfo  {journal} {Nature Nanotechnology}\ }\textbf
  {\bibinfo {volume} {8}},\ \bibinfo {pages} {493} (\bibinfo {year}
  {2013})}\BibitemShut {NoStop}%
\bibitem [{\citenamefont {Reinhardt}\ \emph {et~al.}(2016)\citenamefont
  {Reinhardt}, \citenamefont {Müller}, \citenamefont {Bourassa},\ and\
  \citenamefont {Sankey}}]{Reinhardt2016}%
  \BibitemOpen
  \bibfield  {author} {\bibinfo {author} {\bibfnamefont {C.}~\bibnamefont
  {Reinhardt}}, \bibinfo {author} {\bibfnamefont {T.}~\bibnamefont {Müller}},
  \bibinfo {author} {\bibfnamefont {A.}~\bibnamefont {Bourassa}}, \ and\
  \bibinfo {author} {\bibfnamefont {J.}~\bibnamefont {Sankey}},\ }\href
  {\doibase 10.1103/PhysRevX.6.021001} {\bibfield  {journal} {\bibinfo
  {journal} {Physical Review X}\ }\textbf {\bibinfo {volume} {6}} (\bibinfo
  {year} {2016}),\ 10.1103/PhysRevX.6.021001}\BibitemShut {NoStop}%
\bibitem [{\citenamefont {Saulson}(1990)}]{Saulson19902437}%
  \BibitemOpen
  \bibfield  {author} {\bibinfo {author} {\bibfnamefont {P.}~\bibnamefont
  {Saulson}},\ }\href {\doibase 10.1103/PhysRevD.42.2437} {\bibfield  {journal}
  {\bibinfo  {journal} {Physical Review D}\ }\textbf {\bibinfo {volume} {42}},\
  \bibinfo {pages} {2437} (\bibinfo {year} {1990})}\BibitemShut {NoStop}%
\bibitem [{\citenamefont {Ghadimi}\ \emph {et~al.}(2018)\citenamefont
  {Ghadimi}, \citenamefont {Fedorov}, \citenamefont {Engelsen}, \citenamefont
  {Bereyhi}, \citenamefont {Schilling}, \citenamefont {Wilson},\ and\
  \citenamefont {Kippenberg}}]{Ghadimi2018764}%
  \BibitemOpen
  \bibfield  {author} {\bibinfo {author} {\bibfnamefont {A.}~\bibnamefont
  {Ghadimi}}, \bibinfo {author} {\bibfnamefont {S.}~\bibnamefont {Fedorov}},
  \bibinfo {author} {\bibfnamefont {N.}~\bibnamefont {Engelsen}}, \bibinfo
  {author} {\bibfnamefont {M.}~\bibnamefont {Bereyhi}}, \bibinfo {author}
  {\bibfnamefont {R.}~\bibnamefont {Schilling}}, \bibinfo {author}
  {\bibfnamefont {D.}~\bibnamefont {Wilson}}, \ and\ \bibinfo {author}
  {\bibfnamefont {T.}~\bibnamefont {Kippenberg}},\ }\href {\doibase
  10.1126/science.aar6939} {\bibfield  {journal} {\bibinfo  {journal}
  {Science}\ }\textbf {\bibinfo {volume} {360}},\ \bibinfo {pages} {764}
  (\bibinfo {year} {2018})}\BibitemShut {NoStop}%
\bibitem [{\citenamefont {Tsaturyan}\ \emph {et~al.}(2017)\citenamefont
  {Tsaturyan}, \citenamefont {Barg}, \citenamefont {Polzik},\ and\
  \citenamefont {Schliesser}}]{Tsaturyan2017776}%
  \BibitemOpen
  \bibfield  {author} {\bibinfo {author} {\bibfnamefont {Y.}~\bibnamefont
  {Tsaturyan}}, \bibinfo {author} {\bibfnamefont {A.}~\bibnamefont {Barg}},
  \bibinfo {author} {\bibfnamefont {E.}~\bibnamefont {Polzik}}, \ and\ \bibinfo
  {author} {\bibfnamefont {A.}~\bibnamefont {Schliesser}},\ }\href {\doibase
  10.1038/nnano.2017.101} {\bibfield  {journal} {\bibinfo  {journal} {Nature
  Nanotechnology}\ }\textbf {\bibinfo {volume} {12}},\ \bibinfo {pages} {776}
  (\bibinfo {year} {2017})}\BibitemShut {NoStop}%
\bibitem [{\citenamefont {Vinante}\ \emph {et~al.}(2008)\citenamefont
  {Vinante}, \citenamefont {Bignotto}, \citenamefont {Bonaldi}, \citenamefont
  {Cerdonio}, \citenamefont {Conti}, \citenamefont {Falferi}, \citenamefont
  {Liguori}, \citenamefont {Longo}, \citenamefont {Mezzena}, \citenamefont
  {Ortolan}, \citenamefont {Prodi}, \citenamefont {Salemi}, \citenamefont
  {Taffarello}, \citenamefont {Vedovato}, \citenamefont {Vitale},\ and\
  \citenamefont {Zendri}}]{Vinante2008}%
  \BibitemOpen
  \bibfield  {author} {\bibinfo {author} {\bibfnamefont {A.}~\bibnamefont
  {Vinante}}, \bibinfo {author} {\bibfnamefont {M.}~\bibnamefont {Bignotto}},
  \bibinfo {author} {\bibfnamefont {M.}~\bibnamefont {Bonaldi}}, \bibinfo
  {author} {\bibfnamefont {M.}~\bibnamefont {Cerdonio}}, \bibinfo {author}
  {\bibfnamefont {L.}~\bibnamefont {Conti}}, \bibinfo {author} {\bibfnamefont
  {P.}~\bibnamefont {Falferi}}, \bibinfo {author} {\bibfnamefont
  {N.}~\bibnamefont {Liguori}}, \bibinfo {author} {\bibfnamefont
  {S.}~\bibnamefont {Longo}}, \bibinfo {author} {\bibfnamefont
  {R.}~\bibnamefont {Mezzena}}, \bibinfo {author} {\bibfnamefont
  {A.}~\bibnamefont {Ortolan}}, \bibinfo {author} {\bibfnamefont
  {G.}~\bibnamefont {Prodi}}, \bibinfo {author} {\bibfnamefont
  {F.}~\bibnamefont {Salemi}}, \bibinfo {author} {\bibfnamefont
  {L.}~\bibnamefont {Taffarello}}, \bibinfo {author} {\bibfnamefont
  {G.}~\bibnamefont {Vedovato}}, \bibinfo {author} {\bibfnamefont
  {S.}~\bibnamefont {Vitale}}, \ and\ \bibinfo {author} {\bibfnamefont {J.-P.}\
  \bibnamefont {Zendri}},\ }\href {\doibase 10.1103/PhysRevLett.101.033601}
  {\bibfield  {journal} {\bibinfo  {journal} {Physical Review Letters}\
  }\textbf {\bibinfo {volume} {101}} (\bibinfo {year} {2008}),\
  10.1103/PhysRevLett.101.033601}\BibitemShut {NoStop}%
\bibitem [{\citenamefont {Mertz}, \citenamefont {Marti},\ and\ \citenamefont
  {Mlynek}(1993)}]{Mertz19932344}%
  \BibitemOpen
  \bibfield  {author} {\bibinfo {author} {\bibfnamefont {J.}~\bibnamefont
  {Mertz}}, \bibinfo {author} {\bibfnamefont {O.}~\bibnamefont {Marti}}, \ and\
  \bibinfo {author} {\bibfnamefont {J.}~\bibnamefont {Mlynek}},\ }\href
  {\doibase 10.1063/1.109413} {\bibfield  {journal} {\bibinfo  {journal}
  {Applied Physics Letters}\ }\textbf {\bibinfo {volume} {62}},\ \bibinfo
  {pages} {2344} (\bibinfo {year} {1993})}\BibitemShut {NoStop}%
\bibitem [{\citenamefont {Aspelmeyer}, \citenamefont {Kippenberg},\ and\
  \citenamefont {Marquardt}(2014)}]{Aspelmeyer20141391}%
  \BibitemOpen
  \bibfield  {author} {\bibinfo {author} {\bibfnamefont {M.}~\bibnamefont
  {Aspelmeyer}}, \bibinfo {author} {\bibfnamefont {T.}~\bibnamefont
  {Kippenberg}}, \ and\ \bibinfo {author} {\bibfnamefont {F.}~\bibnamefont
  {Marquardt}},\ }\href {\doibase 10.1103/RevModPhys.86.1391} {\bibfield
  {journal} {\bibinfo  {journal} {Reviews of Modern Physics}\ }\textbf
  {\bibinfo {volume} {86}},\ \bibinfo {pages} {1391} (\bibinfo {year}
  {2014})}\BibitemShut {NoStop}%
\bibitem [{\citenamefont {Serra}\ \emph {et~al.}(2016)\citenamefont {Serra},
  \citenamefont {Bawaj}, \citenamefont {Borrielli}, \citenamefont
  {Di~Giuseppe}, \citenamefont {Forte}, \citenamefont {Kralj}, \citenamefont
  {Malossi}, \citenamefont {Marconi}, \citenamefont {Marin}, \citenamefont
  {Marino}, \citenamefont {Morana}, \citenamefont {Natali}, \citenamefont
  {Pandraud}, \citenamefont {Pontin}, \citenamefont {Prodi}, \citenamefont
  {Rossi}, \citenamefont {Sarro}, \citenamefont {Vitali},\ and\ \citenamefont
  {Bonaldi}}]{Serra2016}%
  \BibitemOpen
  \bibfield  {author} {\bibinfo {author} {\bibfnamefont {E.}~\bibnamefont
  {Serra}}, \bibinfo {author} {\bibfnamefont {M.}~\bibnamefont {Bawaj}},
  \bibinfo {author} {\bibfnamefont {A.}~\bibnamefont {Borrielli}}, \bibinfo
  {author} {\bibfnamefont {G.}~\bibnamefont {Di~Giuseppe}}, \bibinfo {author}
  {\bibfnamefont {S.}~\bibnamefont {Forte}}, \bibinfo {author} {\bibfnamefont
  {N.}~\bibnamefont {Kralj}}, \bibinfo {author} {\bibfnamefont
  {N.}~\bibnamefont {Malossi}}, \bibinfo {author} {\bibfnamefont
  {L.}~\bibnamefont {Marconi}}, \bibinfo {author} {\bibfnamefont
  {F.}~\bibnamefont {Marin}}, \bibinfo {author} {\bibfnamefont
  {F.}~\bibnamefont {Marino}}, \bibinfo {author} {\bibfnamefont
  {B.}~\bibnamefont {Morana}}, \bibinfo {author} {\bibfnamefont
  {R.}~\bibnamefont {Natali}}, \bibinfo {author} {\bibfnamefont
  {G.}~\bibnamefont {Pandraud}}, \bibinfo {author} {\bibfnamefont
  {A.}~\bibnamefont {Pontin}}, \bibinfo {author} {\bibfnamefont
  {G.}~\bibnamefont {Prodi}}, \bibinfo {author} {\bibfnamefont
  {M.}~\bibnamefont {Rossi}}, \bibinfo {author} {\bibfnamefont
  {P.}~\bibnamefont {Sarro}}, \bibinfo {author} {\bibfnamefont
  {D.}~\bibnamefont {Vitali}}, \ and\ \bibinfo {author} {\bibfnamefont
  {M.}~\bibnamefont {Bonaldi}},\ }\href {\doibase 10.1063/1.4953805} {\bibfield
   {journal} {\bibinfo  {journal} {AIP Advances}\ }\textbf {\bibinfo {volume}
  {6}} (\bibinfo {year} {2016}),\ 10.1063/1.4953805}\BibitemShut {NoStop}%
\bibitem [{\citenamefont {Rossi}\ \emph {et~al.}(2017)\citenamefont {Rossi},
  \citenamefont {Kralj}, \citenamefont {Zippilli}, \citenamefont {Natali},
  \citenamefont {Borrielli}, \citenamefont {Pandraud}, \citenamefont {Serra},
  \citenamefont {Di~Giuseppe},\ and\ \citenamefont {Vitali}}]{Rossi2017}%
  \BibitemOpen
  \bibfield  {author} {\bibinfo {author} {\bibfnamefont {M.}~\bibnamefont
  {Rossi}}, \bibinfo {author} {\bibfnamefont {N.}~\bibnamefont {Kralj}},
  \bibinfo {author} {\bibfnamefont {S.}~\bibnamefont {Zippilli}}, \bibinfo
  {author} {\bibfnamefont {R.}~\bibnamefont {Natali}}, \bibinfo {author}
  {\bibfnamefont {A.}~\bibnamefont {Borrielli}}, \bibinfo {author}
  {\bibfnamefont {G.}~\bibnamefont {Pandraud}}, \bibinfo {author}
  {\bibfnamefont {E.}~\bibnamefont {Serra}}, \bibinfo {author} {\bibfnamefont
  {G.}~\bibnamefont {Di~Giuseppe}}, \ and\ \bibinfo {author} {\bibfnamefont
  {D.}~\bibnamefont {Vitali}},\ }\href {\doibase
  10.1103/PhysRevLett.119.123603} {\bibfield  {journal} {\bibinfo  {journal}
  {Physical Review Letters}\ }\textbf {\bibinfo {volume} {119}} (\bibinfo
  {year} {2017}),\ 10.1103/PhysRevLett.119.123603}\BibitemShut {NoStop}%
\bibitem [{\citenamefont {Patil}\ \emph {et~al.}(2015)\citenamefont {Patil},
  \citenamefont {Chakram}, \citenamefont {Chang},\ and\ \citenamefont
  {Vengalattore}}]{Patil2015}%
  \BibitemOpen
  \bibfield  {author} {\bibinfo {author} {\bibfnamefont {Y.}~\bibnamefont
  {Patil}}, \bibinfo {author} {\bibfnamefont {S.}~\bibnamefont {Chakram}},
  \bibinfo {author} {\bibfnamefont {L.}~\bibnamefont {Chang}}, \ and\ \bibinfo
  {author} {\bibfnamefont {M.}~\bibnamefont {Vengalattore}},\ }\href {\doibase
  10.1103/PhysRevLett.115.017202} {\bibfield  {journal} {\bibinfo  {journal}
  {Physical Review Letters}\ }\textbf {\bibinfo {volume} {115}} (\bibinfo
  {year} {2015}),\ 10.1103/PhysRevLett.115.017202},\ \bibinfo {note} {cited By
  40}\BibitemShut {NoStop}%
\bibitem [{\citenamefont {Chowdhury}\ \emph {et~al.}(2019)\citenamefont
  {Chowdhury}, \citenamefont {Vezio}, \citenamefont {Bonaldi}, \citenamefont
  {Borrielli}, \citenamefont {Marino}, \citenamefont {Morana}, \citenamefont
  {Pandraud}, \citenamefont {Pontin}, \citenamefont {Prodi}, \citenamefont
  {Sarro}, \citenamefont {Serra},\ and\ \citenamefont {Marin}}]{Chowdhury2019}%
  \BibitemOpen
  \bibfield  {author} {\bibinfo {author} {\bibfnamefont {A.}~\bibnamefont
  {Chowdhury}}, \bibinfo {author} {\bibfnamefont {P.}~\bibnamefont {Vezio}},
  \bibinfo {author} {\bibfnamefont {M.}~\bibnamefont {Bonaldi}}, \bibinfo
  {author} {\bibfnamefont {A.}~\bibnamefont {Borrielli}}, \bibinfo {author}
  {\bibfnamefont {F.}~\bibnamefont {Marino}}, \bibinfo {author} {\bibfnamefont
  {B.}~\bibnamefont {Morana}}, \bibinfo {author} {\bibfnamefont
  {G.}~\bibnamefont {Pandraud}}, \bibinfo {author} {\bibfnamefont
  {A.}~\bibnamefont {Pontin}}, \bibinfo {author} {\bibfnamefont
  {G.}~\bibnamefont {Prodi}}, \bibinfo {author} {\bibfnamefont
  {P.}~\bibnamefont {Sarro}}, \bibinfo {author} {\bibfnamefont
  {E.}~\bibnamefont {Serra}}, \ and\ \bibinfo {author} {\bibfnamefont
  {F.}~\bibnamefont {Marin}},\ }\href {\doibase 10.1088/2058-9565/ab05f1}
  {\bibfield  {journal} {\bibinfo  {journal} {Quantum Science and Technology}\
  }\textbf {\bibinfo {volume} {4}} (\bibinfo {year} {2019}),\
  10.1088/2058-9565/ab05f1}\BibitemShut {NoStop}%
\bibitem [{\citenamefont {Vezio}\ \emph {et~al.}(2020)\citenamefont {Vezio},
  \citenamefont {Chowdhury}, \citenamefont {Bonaldi}, \citenamefont
  {Borrielli}, \citenamefont {Marino}, \citenamefont {Morana}, \citenamefont
  {Prodi}, \citenamefont {Sarro}, \citenamefont {Serra},\ and\ \citenamefont
  {Marin}}]{Marino2020}%
  \BibitemOpen
  \bibfield  {author} {\bibinfo {author} {\bibfnamefont {P.}~\bibnamefont
  {Vezio}}, \bibinfo {author} {\bibfnamefont {A.}~\bibnamefont {Chowdhury}},
  \bibinfo {author} {\bibfnamefont {M.}~\bibnamefont {Bonaldi}}, \bibinfo
  {author} {\bibfnamefont {A.}~\bibnamefont {Borrielli}}, \bibinfo {author}
  {\bibfnamefont {F.}~\bibnamefont {Marino}}, \bibinfo {author} {\bibfnamefont
  {B.}~\bibnamefont {Morana}}, \bibinfo {author} {\bibfnamefont {G.~A.}\
  \bibnamefont {Prodi}}, \bibinfo {author} {\bibfnamefont {P.~M.}\ \bibnamefont
  {Sarro}}, \bibinfo {author} {\bibfnamefont {E.}~\bibnamefont {Serra}}, \ and\
  \bibinfo {author} {\bibfnamefont {F.}~\bibnamefont {Marin}},\ }\href@noop {}
  {\bibfield  {journal} {\bibinfo  {journal} {Phys. Rev. A}\ }\textbf {\bibinfo
  {volume} {102}},\ \bibinfo {pages} {053505} (\bibinfo {year}
  {2020})}\BibitemShut {NoStop}%
\bibitem [{\citenamefont {Chowdhury}\ \emph {et~al.}(2020)\citenamefont
  {Chowdhury}, \citenamefont {Vezio}, \citenamefont {Bonaldi}, \citenamefont
  {Borrielli}, \citenamefont {Marino}, \citenamefont {Morana}, \citenamefont
  {Prodi}, \citenamefont {Sarro}, \citenamefont {Serra},\ and\ \citenamefont
  {Marin}}]{Chowdhury2020}%
  \BibitemOpen
  \bibfield  {author} {\bibinfo {author} {\bibfnamefont {A.}~\bibnamefont
  {Chowdhury}}, \bibinfo {author} {\bibfnamefont {P.}~\bibnamefont {Vezio}},
  \bibinfo {author} {\bibfnamefont {M.}~\bibnamefont {Bonaldi}}, \bibinfo
  {author} {\bibfnamefont {A.}~\bibnamefont {Borrielli}}, \bibinfo {author}
  {\bibfnamefont {F.}~\bibnamefont {Marino}}, \bibinfo {author} {\bibfnamefont
  {B.}~\bibnamefont {Morana}}, \bibinfo {author} {\bibfnamefont
  {G.}~\bibnamefont {Prodi}}, \bibinfo {author} {\bibfnamefont
  {P.}~\bibnamefont {Sarro}}, \bibinfo {author} {\bibfnamefont
  {E.}~\bibnamefont {Serra}}, \ and\ \bibinfo {author} {\bibfnamefont
  {F.}~\bibnamefont {Marin}},\ }\href {\doibase 10.1103/PhysRevLett.124.023601}
  {\bibfield  {journal} {\bibinfo  {journal} {Physical Review Letters}\
  }\textbf {\bibinfo {volume} {124}} (\bibinfo {year} {2020}),\
  10.1103/PhysRevLett.124.023601}\BibitemShut {NoStop}%
\bibitem [{\citenamefont {Poggio}\ \emph {et~al.}(2007)\citenamefont {Poggio},
  \citenamefont {Degen}, \citenamefont {Mamin},\ and\ \citenamefont
  {Rugar}}]{Poggio2007}%
  \BibitemOpen
  \bibfield  {author} {\bibinfo {author} {\bibfnamefont {M.}~\bibnamefont
  {Poggio}}, \bibinfo {author} {\bibfnamefont {C.}~\bibnamefont {Degen}},
  \bibinfo {author} {\bibfnamefont {H.}~\bibnamefont {Mamin}}, \ and\ \bibinfo
  {author} {\bibfnamefont {D.}~\bibnamefont {Rugar}},\ }\href {\doibase
  10.1103/PhysRevLett.99.017201} {\bibfield  {journal} {\bibinfo  {journal}
  {Physical Review Letters}\ }\textbf {\bibinfo {volume} {99}} (\bibinfo {year}
  {2007}),\ 10.1103/PhysRevLett.99.017201}\BibitemShut {NoStop}%
\bibitem [{not()}]{nota1}%
  \BibitemOpen
  \href@noop {} {\ }\bibinfo {note} {The mechanical dissipation of a resonator
  can be alternatively described by means of a complex spring constant
  $k(1+i\phi(\omega))$, in order to take into account the phase lag
  $\phi(\omega)$ of the displacement behind a sinusoidal force. This damping
  model is more appropriate to describe resonators under low-loss intrinsic
  dissipation as that employed in the paper. However, as we are only interested
  in the spectral analysis of the motion at frequencies close to a mechanical
  resonance, both descriptions are equivalent. Due to the simpler physical
  interpretation, in the rest of the paper we will use the velocity-damping
  model. \cite{Saulson19902437}}\BibitemShut {NoStop}%
\bibitem [{\citenamefont {Kleckner}\ and\ \citenamefont
  {Bouwmeester}(2006)}]{Kleckner200675}%
  \BibitemOpen
  \bibfield  {author} {\bibinfo {author} {\bibfnamefont {D.}~\bibnamefont
  {Kleckner}}\ and\ \bibinfo {author} {\bibfnamefont {D.}~\bibnamefont
  {Bouwmeester}},\ }\href {\doibase 10.1038/nature05231} {\bibfield  {journal}
  {\bibinfo  {journal} {Nature}\ }\textbf {\bibinfo {volume} {444}},\ \bibinfo
  {pages} {75} (\bibinfo {year} {2006})}\BibitemShut {NoStop}%
\bibitem [{\citenamefont {Miller}\ \emph {et~al.}(2018)\citenamefont {Miller},
  \citenamefont {Ansari}, \citenamefont {Heinz}, \citenamefont {Chen},
  \citenamefont {Flader}, \citenamefont {Shin}, \citenamefont {Villanueva},\
  and\ \citenamefont {Kenny}}]{Miller2018}%
  \BibitemOpen
  \bibfield  {author} {\bibinfo {author} {\bibfnamefont {J.}~\bibnamefont
  {Miller}}, \bibinfo {author} {\bibfnamefont {A.}~\bibnamefont {Ansari}},
  \bibinfo {author} {\bibfnamefont {D.}~\bibnamefont {Heinz}}, \bibinfo
  {author} {\bibfnamefont {Y.}~\bibnamefont {Chen}}, \bibinfo {author}
  {\bibfnamefont {I.}~\bibnamefont {Flader}}, \bibinfo {author} {\bibfnamefont
  {D.}~\bibnamefont {Shin}}, \bibinfo {author} {\bibfnamefont {L.}~\bibnamefont
  {Villanueva}}, \ and\ \bibinfo {author} {\bibfnamefont {T.}~\bibnamefont
  {Kenny}},\ }\href {\doibase 10.1063/1.5027850} {\bibfield  {journal}
  {\bibinfo  {journal} {Applied Physics Reviews}\ }\textbf {\bibinfo {volume}
  {5}} (\bibinfo {year} {2018}),\ 10.1063/1.5027850}\BibitemShut {NoStop}%
\bibitem [{\citenamefont {Hammig}\ and\ \citenamefont
  {Wehe}(2007)}]{Hammig2007352}%
  \BibitemOpen
  \bibfield  {author} {\bibinfo {author} {\bibfnamefont {M.}~\bibnamefont
  {Hammig}}\ and\ \bibinfo {author} {\bibfnamefont {D.}~\bibnamefont {Wehe}},\
  }\href {\doibase 10.1109/JSEN.2006.889212} {\bibfield  {journal} {\bibinfo
  {journal} {IEEE Sensors Journal}\ }\textbf {\bibinfo {volume} {7}},\ \bibinfo
  {pages} {352} (\bibinfo {year} {2007})}\BibitemShut {NoStop}%
\bibitem [{\citenamefont {Lee}\ \emph {et~al.}(2010)\citenamefont {Lee},
  \citenamefont {McRae}, \citenamefont {Harris}, \citenamefont {Knittel},\ and\
  \citenamefont {Bowen}}]{Lee2010}%
  \BibitemOpen
  \bibfield  {author} {\bibinfo {author} {\bibfnamefont {K.}~\bibnamefont
  {Lee}}, \bibinfo {author} {\bibfnamefont {T.}~\bibnamefont {McRae}}, \bibinfo
  {author} {\bibfnamefont {G.}~\bibnamefont {Harris}}, \bibinfo {author}
  {\bibfnamefont {J.}~\bibnamefont {Knittel}}, \ and\ \bibinfo {author}
  {\bibfnamefont {W.}~\bibnamefont {Bowen}},\ }\href {\doibase
  10.1103/PhysRevLett.104.123604} {\bibfield  {journal} {\bibinfo  {journal}
  {Physical Review Letters}\ }\textbf {\bibinfo {volume} {104}} (\bibinfo
  {year} {2010}),\ 10.1103/PhysRevLett.104.123604}\BibitemShut {NoStop}%
\bibitem [{\citenamefont {Lee}\ \emph {et~al.}(2002)\citenamefont {Lee},
  \citenamefont {Howell}, \citenamefont {Raman},\ and\ \citenamefont
  {Reifenberger}}]{Lee20021154091}%
  \BibitemOpen
  \bibfield  {author} {\bibinfo {author} {\bibfnamefont {S.}~\bibnamefont
  {Lee}}, \bibinfo {author} {\bibfnamefont {S.}~\bibnamefont {Howell}},
  \bibinfo {author} {\bibfnamefont {A.}~\bibnamefont {Raman}}, \ and\ \bibinfo
  {author} {\bibfnamefont {R.}~\bibnamefont {Reifenberger}},\ }\href {\doibase
  10.1103/PhysRevB.66.115409} {\bibfield  {journal} {\bibinfo  {journal}
  {Physical Review B - Condensed Matter and Materials Physics}\ }\textbf
  {\bibinfo {volume} {66}},\ \bibinfo {pages} {1154091} (\bibinfo {year}
  {2002})}\BibitemShut {NoStop}%
\bibitem [{\citenamefont {Vinante}\ \emph {et~al.}(2013)\citenamefont
  {Vinante}, \citenamefont {Bonaldi}, \citenamefont {Marin},\ and\
  \citenamefont {Zendri}}]{Vinante2013470}%
  \BibitemOpen
  \bibfield  {author} {\bibinfo {author} {\bibfnamefont {A.}~\bibnamefont
  {Vinante}}, \bibinfo {author} {\bibfnamefont {M.}~\bibnamefont {Bonaldi}},
  \bibinfo {author} {\bibfnamefont {F.}~\bibnamefont {Marin}}, \ and\ \bibinfo
  {author} {\bibfnamefont {J.-P.}\ \bibnamefont {Zendri}},\ }\href {\doibase
  10.1038/nnano.2013.130} {\bibfield  {journal} {\bibinfo  {journal} {Nature
  Nanotechnology}\ }\textbf {\bibinfo {volume} {8}},\ \bibinfo {pages} {470}
  (\bibinfo {year} {2013})}\BibitemShut {NoStop}%
\bibitem [{\citenamefont {Harris}\ \emph {et~al.}(2013)\citenamefont {Harris},
  \citenamefont {McAuslan}, \citenamefont {Stace}, \citenamefont {Doherty},\
  and\ \citenamefont {Bowen}}]{Harris2013}%
  \BibitemOpen
  \bibfield  {author} {\bibinfo {author} {\bibfnamefont {G.}~\bibnamefont
  {Harris}}, \bibinfo {author} {\bibfnamefont {D.}~\bibnamefont {McAuslan}},
  \bibinfo {author} {\bibfnamefont {T.}~\bibnamefont {Stace}}, \bibinfo
  {author} {\bibfnamefont {A.}~\bibnamefont {Doherty}}, \ and\ \bibinfo
  {author} {\bibfnamefont {W.}~\bibnamefont {Bowen}},\ }\href {\doibase
  10.1103/PhysRevLett.111.103603} {\bibfield  {journal} {\bibinfo  {journal}
  {Physical Review Letters}\ }\textbf {\bibinfo {volume} {111}} (\bibinfo
  {year} {2013}),\ 10.1103/PhysRevLett.111.103603}\BibitemShut {NoStop}%
\bibitem [{\citenamefont {Gavartin}, \citenamefont {Verlot},\ and\
  \citenamefont {Kippenberg}(2013)}]{Gavartin2013}%
  \BibitemOpen
  \bibfield  {author} {\bibinfo {author} {\bibfnamefont {E.}~\bibnamefont
  {Gavartin}}, \bibinfo {author} {\bibfnamefont {P.}~\bibnamefont {Verlot}}, \
  and\ \bibinfo {author} {\bibfnamefont {T.}~\bibnamefont {Kippenberg}},\
  }\href {\doibase 10.1038/ncomms3860} {\bibfield  {journal} {\bibinfo
  {journal} {Nature Communications}\ }\textbf {\bibinfo {volume} {4}} (\bibinfo
  {year} {2013}),\ 10.1038/ncomms3860}\BibitemShut {NoStop}%
\bibitem [{\citenamefont {Pontin}\ \emph {et~al.}(2014)\citenamefont {Pontin},
  \citenamefont {Bonaldi}, \citenamefont {Borrielli}, \citenamefont
  {Cataliotti}, \citenamefont {Marino}, \citenamefont {Prodi}, \citenamefont
  {Serra},\ and\ \citenamefont {Marin}}]{Pontin2014}%
  \BibitemOpen
  \bibfield  {author} {\bibinfo {author} {\bibfnamefont {A.}~\bibnamefont
  {Pontin}}, \bibinfo {author} {\bibfnamefont {M.}~\bibnamefont {Bonaldi}},
  \bibinfo {author} {\bibfnamefont {A.}~\bibnamefont {Borrielli}}, \bibinfo
  {author} {\bibfnamefont {F.}~\bibnamefont {Cataliotti}}, \bibinfo {author}
  {\bibfnamefont {F.}~\bibnamefont {Marino}}, \bibinfo {author} {\bibfnamefont
  {G.}~\bibnamefont {Prodi}}, \bibinfo {author} {\bibfnamefont
  {E.}~\bibnamefont {Serra}}, \ and\ \bibinfo {author} {\bibfnamefont
  {F.}~\bibnamefont {Marin}},\ }\href {\doibase 10.1103/PhysRevA.89.023848}
  {\bibfield  {journal} {\bibinfo  {journal} {Physical Review A - Atomic,
  Molecular, and Optical Physics}\ }\textbf {\bibinfo {volume} {89}} (\bibinfo
  {year} {2014}),\ 10.1103/PhysRevA.89.023848}\BibitemShut {NoStop}%
\bibitem [{\citenamefont {González}(2000)}]{Gonzalez2000}%
  \BibitemOpen
  \bibfield  {author} {\bibinfo {author} {\bibfnamefont {G.}~\bibnamefont
  {González}},\ }\href {\doibase 10.1088/0264-9381/17/21/305} {\bibfield
  {journal} {\bibinfo  {journal} {Classical and Quantum Gravity}\ }\textbf
  {\bibinfo {volume} {17}},\ \bibinfo {pages} {4409} (\bibinfo {year}
  {2000})}\BibitemShut {NoStop}%
\bibitem [{\citenamefont {Schmid}\ \emph {et~al.}(2011)\citenamefont {Schmid},
  \citenamefont {Jensen}, \citenamefont {Nielsen},\ and\ \citenamefont
  {Boisen}}]{Schmid2011}%
  \BibitemOpen
  \bibfield  {author} {\bibinfo {author} {\bibfnamefont {S.}~\bibnamefont
  {Schmid}}, \bibinfo {author} {\bibfnamefont {K.}~\bibnamefont {Jensen}},
  \bibinfo {author} {\bibfnamefont {K.}~\bibnamefont {Nielsen}}, \ and\
  \bibinfo {author} {\bibfnamefont {A.}~\bibnamefont {Boisen}},\ }\href
  {\doibase 10.1103/PhysRevB.84.165307} {\bibfield  {journal} {\bibinfo
  {journal} {Physical Review B - Condensed Matter and Materials Physics}\
  }\textbf {\bibinfo {volume} {84}} (\bibinfo {year} {2011}),\
  10.1103/PhysRevB.84.165307},\ \bibinfo {note} {cited By 97}\BibitemShut
  {NoStop}%
\bibitem [{\citenamefont {Purdy}\ \emph {et~al.}(2014)\citenamefont {Purdy},
  \citenamefont {Yu}, \citenamefont {Peterson}, \citenamefont {Kampel},\ and\
  \citenamefont {Regal}}]{Purdy2014}%
  \BibitemOpen
  \bibfield  {author} {\bibinfo {author} {\bibfnamefont {T.}~\bibnamefont
  {Purdy}}, \bibinfo {author} {\bibfnamefont {P.-L.}\ \bibnamefont {Yu}},
  \bibinfo {author} {\bibfnamefont {R.}~\bibnamefont {Peterson}}, \bibinfo
  {author} {\bibfnamefont {N.}~\bibnamefont {Kampel}}, \ and\ \bibinfo {author}
  {\bibfnamefont {C.}~\bibnamefont {Regal}},\ }\href {\doibase
  10.1103/PhysRevX.3.031012} {\bibfield  {journal} {\bibinfo  {journal}
  {Physical Review X}\ }\textbf {\bibinfo {volume} {3}} (\bibinfo {year}
  {2014}),\ 10.1103/PhysRevX.3.031012}\BibitemShut {NoStop}%
\bibitem [{\citenamefont {Wilson}\ \emph {et~al.}(2009)\citenamefont {Wilson},
  \citenamefont {Regal}, \citenamefont {Papp},\ and\ \citenamefont
  {Kimble}}]{Wilson2009}%
  \BibitemOpen
  \bibfield  {author} {\bibinfo {author} {\bibfnamefont {D.}~\bibnamefont
  {Wilson}}, \bibinfo {author} {\bibfnamefont {C.}~\bibnamefont {Regal}},
  \bibinfo {author} {\bibfnamefont {S.}~\bibnamefont {Papp}}, \ and\ \bibinfo
  {author} {\bibfnamefont {H.}~\bibnamefont {Kimble}},\ }\href {\doibase
  10.1103/PhysRevLett.103.207204} {\bibfield  {journal} {\bibinfo  {journal}
  {Physical Review Letters}\ }\textbf {\bibinfo {volume} {103}} (\bibinfo
  {year} {2009}),\ 10.1103/PhysRevLett.103.207204}\BibitemShut {NoStop}%
\bibitem [{\citenamefont {Borrielli}\ \emph {et~al.}(2016)\citenamefont
  {Borrielli}, \citenamefont {Marconi}, \citenamefont {Marin}, \citenamefont
  {Marino}, \citenamefont {Morana}, \citenamefont {Pandraud}, \citenamefont
  {Pontin}, \citenamefont {Prodi}, \citenamefont {Sarro}, \citenamefont
  {Serra},\ and\ \citenamefont {Bonaldi}}]{Borrielli2016}%
  \BibitemOpen
  \bibfield  {author} {\bibinfo {author} {\bibfnamefont {A.}~\bibnamefont
  {Borrielli}}, \bibinfo {author} {\bibfnamefont {L.}~\bibnamefont {Marconi}},
  \bibinfo {author} {\bibfnamefont {F.}~\bibnamefont {Marin}}, \bibinfo
  {author} {\bibfnamefont {F.}~\bibnamefont {Marino}}, \bibinfo {author}
  {\bibfnamefont {B.}~\bibnamefont {Morana}}, \bibinfo {author} {\bibfnamefont
  {G.}~\bibnamefont {Pandraud}}, \bibinfo {author} {\bibfnamefont
  {A.}~\bibnamefont {Pontin}}, \bibinfo {author} {\bibfnamefont
  {G.}~\bibnamefont {Prodi}}, \bibinfo {author} {\bibfnamefont
  {P.}~\bibnamefont {Sarro}}, \bibinfo {author} {\bibfnamefont
  {E.}~\bibnamefont {Serra}}, \ and\ \bibinfo {author} {\bibfnamefont
  {M.}~\bibnamefont {Bonaldi}},\ }\href {\doibase 10.1103/PhysRevB.94.121403}
  {\bibfield  {journal} {\bibinfo  {journal} {Physical Review B}\ }\textbf
  {\bibinfo {volume} {94}} (\bibinfo {year} {2016}),\
  10.1103/PhysRevB.94.121403}\BibitemShut {NoStop}%
\bibitem [{\citenamefont {Serra}\ \emph {et~al.}(2018)\citenamefont {Serra},
  \citenamefont {Morana}, \citenamefont {Borrielli}, \citenamefont {Marin},
  \citenamefont {Pandraud}, \citenamefont {Pontin}, \citenamefont {Prodi},
  \citenamefont {Sarro},\ and\ \citenamefont {Bonaldi}}]{Serra20181193}%
  \BibitemOpen
  \bibfield  {author} {\bibinfo {author} {\bibfnamefont {E.}~\bibnamefont
  {Serra}}, \bibinfo {author} {\bibfnamefont {B.}~\bibnamefont {Morana}},
  \bibinfo {author} {\bibfnamefont {A.}~\bibnamefont {Borrielli}}, \bibinfo
  {author} {\bibfnamefont {F.}~\bibnamefont {Marin}}, \bibinfo {author}
  {\bibfnamefont {G.}~\bibnamefont {Pandraud}}, \bibinfo {author}
  {\bibfnamefont {A.}~\bibnamefont {Pontin}}, \bibinfo {author} {\bibfnamefont
  {G.}~\bibnamefont {Prodi}}, \bibinfo {author} {\bibfnamefont
  {P.}~\bibnamefont {Sarro}}, \ and\ \bibinfo {author} {\bibfnamefont
  {M.}~\bibnamefont {Bonaldi}},\ }\href {\doibase 10.1109/JMEMS.2018.2876593}
  {\bibfield  {journal} {\bibinfo  {journal} {Journal of Microelectromechanical
  Systems}\ }\textbf {\bibinfo {volume} {27}},\ \bibinfo {pages} {1193}
  (\bibinfo {year} {2018})}\BibitemShut {NoStop}%
\bibitem [{\citenamefont {Unterreithmeier}, \citenamefont {Weig},\ and\
  \citenamefont {Kotthaus}(2009)}]{Unterreithmeier20091001}%
  \BibitemOpen
  \bibfield  {author} {\bibinfo {author} {\bibfnamefont {Q.}~\bibnamefont
  {Unterreithmeier}}, \bibinfo {author} {\bibfnamefont {E.}~\bibnamefont
  {Weig}}, \ and\ \bibinfo {author} {\bibfnamefont {J.}~\bibnamefont
  {Kotthaus}},\ }\href {\doibase 10.1038/nature07932} {\bibfield  {journal}
  {\bibinfo  {journal} {Nature}\ }\textbf {\bibinfo {volume} {458}},\ \bibinfo
  {pages} {1001} (\bibinfo {year} {2009})}\BibitemShut {NoStop}%
\bibitem [{\citenamefont {Buters}\ \emph {et~al.}(2017)\citenamefont {Buters},
  \citenamefont {Heeck}, \citenamefont {Eerkens}, \citenamefont {Weaver},
  \citenamefont {Luna}, \citenamefont {De~Man},\ and\ \citenamefont
  {Bouwmeester}}]{Buters2017}%
  \BibitemOpen
  \bibfield  {author} {\bibinfo {author} {\bibfnamefont {F.}~\bibnamefont
  {Buters}}, \bibinfo {author} {\bibfnamefont {K.}~\bibnamefont {Heeck}},
  \bibinfo {author} {\bibfnamefont {H.}~\bibnamefont {Eerkens}}, \bibinfo
  {author} {\bibfnamefont {M.}~\bibnamefont {Weaver}}, \bibinfo {author}
  {\bibfnamefont {F.}~\bibnamefont {Luna}}, \bibinfo {author} {\bibfnamefont
  {S.}~\bibnamefont {De~Man}}, \ and\ \bibinfo {author} {\bibfnamefont
  {D.}~\bibnamefont {Bouwmeester}},\ }\href {\doibase 10.1063/1.4978212}
  {\bibfield  {journal} {\bibinfo  {journal} {Applied Physics Letters}\
  }\textbf {\bibinfo {volume} {110}} (\bibinfo {year} {2017}),\
  10.1063/1.4978212}\BibitemShut {NoStop}%
\bibitem [{\citenamefont {Schmid}\ \emph {et~al.}(2014)\citenamefont {Schmid},
  \citenamefont {Bagci}, \citenamefont {Zeuthen}, \citenamefont {Taylor},
  \citenamefont {Herring}, \citenamefont {Cassidy}, \citenamefont {Marcus},
  \citenamefont {Guillermo~Villanueva}, \citenamefont {Amato}, \citenamefont
  {Boisen}, \citenamefont {Shin}, \citenamefont {Kong}, \citenamefont
  {Sørensen}, \citenamefont {Usami},\ and\ \citenamefont
  {Polzik}}]{Schmid2014}%
  \BibitemOpen
  \bibfield  {author} {\bibinfo {author} {\bibfnamefont {S.}~\bibnamefont
  {Schmid}}, \bibinfo {author} {\bibfnamefont {T.}~\bibnamefont {Bagci}},
  \bibinfo {author} {\bibfnamefont {E.}~\bibnamefont {Zeuthen}}, \bibinfo
  {author} {\bibfnamefont {J.}~\bibnamefont {Taylor}}, \bibinfo {author}
  {\bibfnamefont {P.}~\bibnamefont {Herring}}, \bibinfo {author} {\bibfnamefont
  {M.}~\bibnamefont {Cassidy}}, \bibinfo {author} {\bibfnamefont
  {C.}~\bibnamefont {Marcus}}, \bibinfo {author} {\bibfnamefont
  {L.}~\bibnamefont {Guillermo~Villanueva}}, \bibinfo {author} {\bibfnamefont
  {B.}~\bibnamefont {Amato}}, \bibinfo {author} {\bibfnamefont
  {A.}~\bibnamefont {Boisen}}, \bibinfo {author} {\bibfnamefont
  {Y.}~\bibnamefont {Shin}}, \bibinfo {author} {\bibfnamefont {J.}~\bibnamefont
  {Kong}}, \bibinfo {author} {\bibfnamefont {A.}~\bibnamefont {Sørensen}},
  \bibinfo {author} {\bibfnamefont {K.}~\bibnamefont {Usami}}, \ and\ \bibinfo
  {author} {\bibfnamefont {E.}~\bibnamefont {Polzik}},\ }\href {\doibase
  10.1063/1.4862296} {\bibfield  {journal} {\bibinfo  {journal} {Journal of
  Applied Physics}\ }\textbf {\bibinfo {volume} {115}} (\bibinfo {year}
  {2014}),\ 10.1063/1.4862296}\BibitemShut {NoStop}%
\bibitem [{\citenamefont {Krick}, \citenamefont {Lenahan},\ and\ \citenamefont
  {Kanicki}(1988)}]{Krick19888226}%
  \BibitemOpen
  \bibfield  {author} {\bibinfo {author} {\bibfnamefont {D.}~\bibnamefont
  {Krick}}, \bibinfo {author} {\bibfnamefont {P.}~\bibnamefont {Lenahan}}, \
  and\ \bibinfo {author} {\bibfnamefont {J.}~\bibnamefont {Kanicki}},\ }\href
  {\doibase 10.1103/PhysRevB.38.8226} {\bibfield  {journal} {\bibinfo
  {journal} {Physical Review B}\ }\textbf {\bibinfo {volume} {38}},\ \bibinfo
  {pages} {8226} (\bibinfo {year} {1988})}\BibitemShut {NoStop}%
\bibitem [{\citenamefont {Sommer}\ and\ \citenamefont
  {Genes}(2019)}]{Sommer2019}%
  \BibitemOpen
  \bibfield  {author} {\bibinfo {author} {\bibfnamefont {C.}~\bibnamefont
  {Sommer}}\ and\ \bibinfo {author} {\bibfnamefont {C.}~\bibnamefont {Genes}},\
  }\href {\doibase 10.1103/PhysRevLett.123.203605} {\bibfield  {journal}
  {\bibinfo  {journal} {Physical Review Letters}\ }\textbf {\bibinfo {volume}
  {123}} (\bibinfo {year} {2019}),\ 10.1103/PhysRevLett.123.203605}\BibitemShut
  {NoStop}%
\bibitem [{\citenamefont {Thompson}\ \emph {et~al.}(2008)\citenamefont
  {Thompson}, \citenamefont {Zwickl}, \citenamefont {Jayich}, \citenamefont
  {Marquardt}, \citenamefont {Girvin},\ and\ \citenamefont
  {Harris}}]{Thompson200872}%
  \BibitemOpen
  \bibfield  {author} {\bibinfo {author} {\bibfnamefont {J.}~\bibnamefont
  {Thompson}}, \bibinfo {author} {\bibfnamefont {B.}~\bibnamefont {Zwickl}},
  \bibinfo {author} {\bibfnamefont {A.}~\bibnamefont {Jayich}}, \bibinfo
  {author} {\bibfnamefont {F.}~\bibnamefont {Marquardt}}, \bibinfo {author}
  {\bibfnamefont {S.}~\bibnamefont {Girvin}}, \ and\ \bibinfo {author}
  {\bibfnamefont {J.}~\bibnamefont {Harris}},\ }\href {\doibase
  10.1038/nature06715} {\bibfield  {journal} {\bibinfo  {journal} {Nature}\
  }\textbf {\bibinfo {volume} {452}},\ \bibinfo {pages} {72} (\bibinfo {year}
  {2008})},\ \bibinfo {note} {cited By 976}\BibitemShut {NoStop}%
\bibitem [{\citenamefont {Poot}\ and\ \citenamefont {van~der
  Zant}(2012)}]{Poot2012273}%
  \BibitemOpen
  \bibfield  {author} {\bibinfo {author} {\bibfnamefont {M.}~\bibnamefont
  {Poot}}\ and\ \bibinfo {author} {\bibfnamefont {H.}~\bibnamefont {van~der
  Zant}},\ }\href {\doibase 10.1016/j.physrep.2011.12.004} {\bibfield
  {journal} {\bibinfo  {journal} {Physics Reports}\ }\textbf {\bibinfo {volume}
  {511}},\ \bibinfo {pages} {273} (\bibinfo {year} {2012})}\BibitemShut
  {NoStop}%
\bibitem [{\citenamefont {Marquardt}\ and\ \citenamefont
  {Girvin}(2009)}]{Marquardt2009}%
  \BibitemOpen
  \bibfield  {author} {\bibinfo {author} {\bibfnamefont {F.}~\bibnamefont
  {Marquardt}}\ and\ \bibinfo {author} {\bibfnamefont {S.}~\bibnamefont
  {Girvin}},\ }\href {\doibase 10.1103/Physics.2.40} {\bibfield  {journal}
  {\bibinfo  {journal} {Physics}\ }\textbf {\bibinfo {volume} {2}} (\bibinfo
  {year} {2009}),\ 10.1103/Physics.2.40}\BibitemShut {NoStop}%
\end{thebibliography}%

\end{document}